\documentclass[preprint]{aastex}

\slugcomment{Accepted for publication in ApJ, March 2005}

\shorttitle{Optical linear polarization of late M- and L-type dwarfs}
\shortauthors{Zapatero Osorio et al.}

\begin{document}

\small

\title{Optical linear polarization of late M- and L-type dwarfs}

\author{M.\,R$.$ Zapatero Osorio}
\affil{LAEFF-INTA, P.\,O$.$ 50727, E-28080 Madrid, Spain}
\email{mosorio@laeff.esa.es}

\author{J.\,A$.$ Caballero}
\affil{Instituto de Astrof\'\i sica de Canarias, E-38200 La Laguna, 
       Tenerife, Spain}
\email{zvezda@ll.iac.es}

\and

\author{V.\,J.\,S$.$ B\'ejar}
\affil{Instituto de Astrof\'\i sica de Canarias, GTC Project. E-38200 
       La Laguna, Tenerife, Spain}
\email{vbejar@ll.iac.es}


\begin{abstract}
We report on the linear polarimetric observations in the Johnson $I$-band filter of 44 ultracool dwarfs with spectral types between M6 and L7.5, corresponding to effective temperatures in the range 2800--1400\,K, and one M4.5-type star. Based on our measurements of polarization ($P$) and their associated error bars ($\sigma_P$), eleven (10 L and 1 M) dwarfs appear to have significant linear polarization ($P/\sigma_P$\,$\ge$\,3). For these, the polarization degrees we have measured are in the interval $P$\,=\,0.2--2.5\%. Because of the typical average uncertainty of our data, we can easily confirm polarization of ultracool dwarfs that show degree of linear polarization greater than 0.4\%. We have compared the two populations in our sample, the M- and L-type dwarfs, and have found evidence for a larger frequency of high $I$-band polarization in the coolest objects, supporting the presence of significant amounts of dust in L-type atmospheres. The probable mechanism polarizing the far-red optical photons of ultracool dwarfs is related to the presence of heterogeneous dust clouds nonuniformly distributed across the visible photospheres and the asymmetric shape of the objects (rapid rotations impose deviations from sphericity). In some young ultracool dwarfs, surrounding dusty disks (or shells) may also yield polarization. For polarimetric detections, a trend for slightly larger polarization from L0 to L6.5 may be present in our data, suggesting changes in the distribution of the grain properties and in the vertical height of the cloud layer. Faster rotations and important differences in metallicity and age within our sample could also account for this trend. One of the targets is the peculiar brown dwarf 2MASS\,J22443167$+$2043433 (L6.5), for which we have determined the largest $I$-band polarization degree in our study. We discuss that the origin of such large polarization may lie in a surrounding dusty disk (or shell) and/or rather large photospheric dust grains. Two of the likely polarized dwarfs (CFHT-BD-Tau\,4, a very young, M7-type brown dwarf of the Taurus star-forming region, and 2MASS\,J00361617$+$1821104, an L3.5 field dwarf) were also observed in the Johnson $R$-band filter, allowing us to discuss qualitatively the size of the grains responsible for the polarization. Our data support the presence of a circum(sub)stellar disk around the young accreting brown dwarf CFHT-BD-Tau\,4. The higher degree of polarization in the $R$-band as compared to the $I$-band indicates that the grain growth lies in the submicron regime in the visible photosphere of 2MASS\,J00361617$+$1821104 (effective temperature of about 1900\,K). Our polarimetric data do not obviously correlate with activity (H$\alpha$ and radio emission) or projected rotational velocity. Three polarized early- to mid-L dwarfs have been photometrically monitored in the $I$-band, displaying light curves with amplitudes below 10\,mmag.
\end{abstract}

\keywords{polarization --- stars: atmospheres ---  stars: late-type --- stars: low-mass, brown dwarfs 
}


\section{Introduction}

Very low-mass dwarf stars and brown dwarfs are characterized by effective temperatures ($T_{\rm eff}$) below $\sim$2800\,K, corresponding to spectral types later than M6--M7. These objects are often called ``ultracool'' dwarfs in the literature (e.g$.$ Liebert et al$.$ \cite{liebert99}; Gizis et al$.$ \cite{gizis00}). During the last decade, a large population of ultracool dwarfs covering an interval of $T_{\rm eff}$ between 2800 and 800\,K have been discovered by different groups of observers (see the reviews by Basri \cite{basri00b} and Chabrier \& Baraffe \cite{chabrier00}). They have been assigned spectral types late-M, L and T. A detailed description of the features used for spectral classification is provided by the following works: Mart\'\i n et al$.$ \cite{martin99a}; Kirkpatrick et al$.$ \cite{kirk99}; Geballe et al$.$ \cite{geballe02}; and Burgasser et al$.$ \cite{burgasser02a}. Refractory elements, like Ti, Fe, Ca and V, are strongly depleted from the gas in the atmospheres of ultracool dwarfs, forming clouds of dust grains (Tsuji et al$.$ \cite{tsuji96}). Observations of the coolest dwarfs of spectral class T in combination with theory suggest that, below $\sim$1300\,K, grains predominantly lie in a thin deck at the very deep photosphere, i.e$.$ beyond the visible region of the atmosphere, because of complete gravitational settling (Allard et al$.$ \cite{allard01}; Marley et al$.$ \cite{marley02}). However, for the L type, there is inefficient sedimentation, and dust clouds play an important role in controlling opacity and the temperature structure of the atmosphere. The physical properties (e.g$.$ shape, size), the number density and the precise species of the dust particulates remain highly unknown. Also unknown are the geometrical height and the location of the cloud layer within the atmospheres.

Recently, Sengupta \& Krishan \cite{sengupta01} argued that detectable polarization could arise because of dust scattering in the nonspherical atmosphere of L dwarfs. The size of the grains in the upper photosphere is expected to be in the submicron range; hence, polarization should occur at optical wavelengths. Nonzero polarization can also occur in the infrared if the particle size is large. The pioneering work by M\'enard et al$.$ \cite{menard02} confirmed the presence of red optical polarization in L dwarfs. Their work also suggested that a large fraction (about 50\%) of ultracool dwarfs shows significant polarized $I$-band radiation ($I_{\rm Bessel}$ filter centred on 768\,nm). These authors measured degrees of linear polarization of the order of 0.1--0.2\%~in three L-type field dwarfs, and attributed such polarization to the probable presence of dust clouds in the atmosphere of their targets. Sengupta \cite{sengupta03} tried to constrain the size of the grains by fitting models to the $I$-band observations of M\'enard et al$.$ \cite{menard02}.

Under the hypothesis that the linear polarization observed by M\'enard et al$.$ \cite{menard02} is due to dust scattering, there are several factors that make net disk-integrated polarization more likely. First, nonspherical grains in the atmospheres. Second, the lack of symmetry in the shape of the objects, which leads to incomplete cancellation of the polarization of the radiation. The very fast rotation observed in the great majority of the known ultracool dwarfs (e.g$.$ Basri \cite{basri00b}) gives them the shape of an oblate ellipsoid. Third, and of especial interest, dust clouds may evolve rapidly because of intense vertical motions, which are due to convective heat transport from the interior (S\'anchez-Lavega \cite{sanchez01}). Evidence for time evolution of the emergent flux of ultracool dwarfs in intervals of one to several rotation periods is provided by the photometric monitoring carried out by various groups (e.g$.$ Bailer-Jones \& Mundt \cite{bailer01}; Mart\'\i n et al$.$ \cite{martin01a}; Gelino et al$.$ \cite{gelino02}; Zapatero Osorio et al$.$ \cite{osorio03}; see also the discussion by Bailer-Jones \& Lamm \cite{bailer03}). The rapidly evolving heterogeneous distribution of dust clouds may contribute to variations in the degree of polarization. In this respect, measurements of linear polarization can help identify the scattering mechanism, it can pinpoint an obscure source, and it can give information on the physical properties of the grains or the scattering medium. Depending on the geometry, polarimetric monitoring can also shed new light on the ``weather'' of ultracool dwarfs. 

In this paper, we report on the linear polarimetric measurements in the $R$- and $I$-bands of 44 ultracool dwarfs with spectral types in the range M6--L7.5, and one M4.5V dwarf. We detect significant polarized optical radiation in several of them, and discuss measured and zero linear polarization as a function of the known spectroscopic and photometric properties of the target sample.

\section{Target selection \label{target_selection}}

The total of 45 program objects are provided in Table~\ref{observations} ordered by increasing right ascension. The great majority of them were selected from the DENIS and 2MASS near-infrared sky surveys and the optical Sloan survey (``discovery'' papers are the following: Delfosse et al$.$ \cite{delfosse97}; Kirkpatrick et al$.$ \cite{kirk99}, \cite{kirk00}; Reid et al$.$ \cite{reid00}; Gizis et al$.$ \cite{gizis00}; Gizis \cite{gizis02}; Liebert et al$.$ \cite{liebert03}; McLean et al$.$ \cite{mclean03}; Dahn et al$.$ \cite{dahn02}; Hawley et al$.$ \cite{hawley02}; Wilson et al$.$ \cite{wilson03}; Cruz et al$.$ \cite{cruz03}; Kendall et al$.$ \cite{kendall04}). Many (33) have spectral types in the range L0--L7.5V (as measured from optical and near-infrared spectra), with estimated effective temperatures from $\sim$2500\,K down to $\sim$1400\,K (Vrba et al$.$ \cite{vrba04}). Additional program objects with L classes are Kelu\,1 (L2.5V, Ruiz, Leggett, \& Allard \cite{ruiz97}), and the recently discovered low-metallicity L-type dwarf LSR\,J1610$-$0040 (L\'epine et al$.$ \cite{lepine03}). This makes a total of 35 L dwarfs in our study. For comparison purposes, we also observed ten mid-M to late-M dwarfs, including BRI\,0021--0214 (M9.5V, Irwin, McMahon, \& Reid \cite{irwin91}) and the M7-type young brown dwarf CFHT-BD-Tau 4 (Mart\'\i n et al$.$ \cite{martin01b}). From now on, we will use ``common'' abridged names to refer to the various objects. 

All the late-M and L-type targets were selected to be amongst the brightest sources for their classes, and have masses very likely in the interval 0.03--0.15\,$M_{\odot}$, i.e$.$ very low-mass stars and brown dwarfs. They are located at very nearby distances. The astrometric parallaxes of nine dwarfs in our sample (BRI\,0021, J0036$+$18, Kelu\,1, J1439$+$19, J1507$-$16, J1658$+$70, LHS\,3406, GJ\,4281, and J2224$-$01) place them between 7 and 19\,pc (Dahn et al$.$ \cite{dahn02}). On the assumption that the rest of the program dwarfs have similar ages and after comparison of their apparent magnitudes with the mean absolute magnitudes for each spectral type, we deduce that all the targets very likely lie within 35\,pc from the Sun, except for CFHT-BD-Tau 4. This M7-type dwarf is a very young brown dwarf member of the Taurus star-forming region (Mart\'\i n et al$.$ \cite{martin01b}), and is presumably located at a distance of 140\,pc. L\'epine et al$.$ \cite{lepine03} reported an upper limit of 30\,pc on the distance to the metal-depleted dwarf J1610$-$00, being the most likely value around 16\,pc. The stellar polarizations (mostly interstellar in origin) recently found by Weitenbeck \cite{weitenbeck04} correspond to field stars that are beyond 200\,pc. Moreover, the optical study of Tamburini et al$.$ \cite{tamburini02}, based on more than 1000 stars, indicates that the contribution to the polarization by the interstellar medium becomes effective only after $\sim$70\,pc. Except for CFHT-BD-Tau 4, none of our program dwarfs lies at such long distances. Because of their marked proximity, we do not expect significant polarization induced by the interstellar medium, and we will not attempt to remove the interstellar contribution from our polarimetric measurements.

\section{Observations}

Polarimetric data were collected with the Calar Alto Faint Object Spectrograph (CAFOS) instrument attached to the 2.2-m telescope at the Calar Alto Observatory (Spain). In its imaging polarimetric mode, CAFOS, which is mounted at the Cassegrain focus, is equipped with a Wollaston prism with an effective beam separation of 18.5\,arcsec, plus a rotable half-wave retarder plate and a stripe mask. This combination provides the capability to measure linear polarization by means of dual beam imaging polarimetry. CAFOS is also equipped with a SITe 2048$\times$2048 pixels CCD detector (image scale of 0.53\,arcsec/pixel), which was windowed to the central 1024$\times$1024 pixels providing a field of view of 9$\times$9\,arcmin$^2$. The observations were obtained in the following Universal Time (UT) dates corresponding to three different runs: 2003 Aug 28--31 and Sep 02, 2004 May 16--21, and 2004 Aug 12--19. Except for 2003 Sep 02, the remaining 2003 nights and the 2004 Aug nights were photometric; the 2004 May run was hampered by thin cirrus and high relative humidity.

Images were obtained in the Johnson $R$- and $I$-band filters centred on 641 and 850\,nm, respectively. The passband of these filters is 158 ($R$) and 150\,nm ($I$). All program objects were observed in the $I$-band because they are relatively brighter at these wavelengths than at shorter wavelengths, and only two of the targets were observed in the $R$-band (CFHT-BD-Tau\,4 and J0036$+$18). Raw data were bias subtracted and flat-fielded using twilight flat frames, which were taken through each of the broad-band filters and without the polarizing optics. The complete set of calibration images, i.e$.$ bias and twilight flat frames, were collected every night. We did not detect any systematic variation of these images from night to night. Both program sources and standard stars were observed at the same spot within the detector (around pixel 512, 512 of the windowed frames), which is very close to the optical axis of the telescope/instrument system. For linear polarimetry, we collected images at four different angles of the half-wave retarder plate: 0, 45, 22.5 and 67.5\,deg. We provide in Table~\ref{observations} the UT date of the observations, the exposure time and number of images obtained per position of the retarder plate, and the full-width-half-maximum (FWHM) and air mass at the time of the observations. The FWHM as measured on the reduced images ranged from 1.0 to 2.9\,arcsec.

For the calibration of the polarimetric measurements we observed four high-polarization standard stars, namely Hiltner 960, BD$+$64 106 (Schmidt, Elston, \& Lupie \cite{schmidt92}), Cyg\,OB2 7 and Cyg\,OB2 17 (Whittet et al$.$ \cite{whittet92}), on 30 different occasions. These stars show linear polarization degrees between 3.0\%~and 5.2\%~in the $R$- and $I$-bands. The two Cyg\,OB2 standards were observed only in $I$ in the 2004 May campaing, while Hiltner 960 and BD$+$64 106 were observed in the two filters on many occasions during the three campaings. This has allowed us to check for the efficiency and stability of the instrument from night to night and from run to run. Our linear polarization measurements of the reference stars have typical standard deviations of $\sigma_P$\,=\,0.03--0.10\%~(depending on the brightness of the source), and appear to be systematically smaller than the catalog values by a factor of 1.071\,$\pm$\,0.004 ($I$-band). This indicates that an efficiency correction might be needed. However, the great majority of our program objects does not show significant linear polarization. If some polarization is detected, it is below $P$\,=\,2.5\%, and the rectification for the instrumental efficiency loss is of the order of (or even smaller than) the error bar of the measurement. Thus, we have not applied any correction to our data. Regarding the position angle of the polarization vibration, $\theta$, we have determined that there is a zero-point correction, which is $-$3.2$\pm$1.1\,deg. The instrumental polarization was also checked by measuring two unpolarized standard stars, GD\,319 and BD$+$28 4211 (Turnshek et al$.$ \cite{turnshek90}), on 11 different occasions during the three campaings. All the measurements are compatible with zero polarization within 1-$\sigma$ the error bars, suggesting that the position within the detector at which all sources were observed is free of instrumental polarization.

\section{Linear polarimetric results}

We have obtained the linear polarization degrees and the polarization angles by calculating the normalized Stokes parameters, $Q/I$ and $U/I$, from the recorded images. These quantities depend on ratios of fluxes at one filter; the $Q/I$ parameter is derived from the pair of exposures with the half-wave plate at 0\,deg and 45\,deg, and $U/I$ is computed from the images with the half-wave plate at 22.5\,deg and 67.5\,deg. The mathematical expressions used are as follows:
\begin{equation}
R^2_Q\,=\,\frac{o(0^\circ)/e(0^\circ)}{o(45^\circ)/e(45^\circ)}; 
~~~~Q/I\,=\,\frac{R_Q-1}{R_Q+1}
\end{equation}
\begin{equation}
R^2_U\,=\,\frac{o(22.5^\circ)/e(22.5^\circ)}{o(67.5^\circ)/e(67.5^\circ)}; 
~~~~U/I\,=\,\frac{R_U-1}{R_U+1}
\end{equation}
\begin{equation}
P\,=\,\sqrt{(Q/I)^2\,+\,(U/I)^2}
\end{equation}
\begin{equation}
\theta\,=\,0.5\,{\rm tan^{-1}}\left(\frac{U/I}{Q/I}\right)
\end{equation}
where $o$ (ordinary) and $e$ (extraordinary) refer to the fluxes of the dual images of the program source on a single frame, and $P$ and $\theta$ are the linear polarization and the polarization angle, respectively. The quantities $R_Q$ and $R_U$, used to evaluate the normalized Stokes parameters, correct for possible flat flaws. Fluxes have been derived using different apertures around the targets: 0.5, 0.8, 1.0, 1.2 and 1.5 times the average FWHM of each frame. Larger apertures have not been considered to avoid contamination from nearby sources and variable sky contributions. We have finally chosen the 1.0$\times$FWHM aperture because it both minimized the photon contribution of nearby contaminants and maximized the signal-to-noise ratio of the measurements. In case of very good seeing (FWHM\,$\le$\,1.3\,arcsec), the best aperture was 1.2$\times$FWHM. We provide our results in Table~\ref{polarimetry1}, where the names of the program objects have been abreviated conveniently. The measured Stokes parameters and the degree of linear polarization are listed together with the filter and Modified Julian Date (MJD) of the observations. Error bars are determined from the standard deviation of the multiple measurements (the great majority of the targets were observed more than once in each position of the half-wave retarder). If only one measure is available, the uncertainty is estimated from the various apertures. Whenever there are more than one epoch of observations, e.g$.$ CFHT-BD-Tau\,4, which was observed on four different nights, we provide the average of all the individual measurements together with the error of the mean in Table~\ref{polarimetry2}. Uncertainties in position angle are not well-defined for the unpolarized targets, and thus are not listed in Tables~\ref{polarimetry1} and \ref{polarimetry2}. Note that in Tables~\ref{polarimetry1}--\ref{polarimetry3} all values of $Q/I$, $U/I$ and $P$ are quoted as measured, i.e$.$ without any correction for instrumental efficiency loss or correction for the statistical bias which affects polarimetry at small values of polarimetric signal-to-noise ratio (e.g$.$ Simmons \& Stewart \cite{simmons85}). 

To select the likely polarized sources amongst our targets, we have applied the following criterion: the derived degree of polarization must exceed three times the associated uncertainty ($P/\sigma_P$\,$\ge$\,3). On the assumption of a Gaussian distribution of the measurements within their error bars, such a criterion sets the confidence of positive detections at the level of 99\%. If $U/I$ is lacking, we have used $Q/I$ as an indicator for the presence of some polarization. In our sample, eleven objects appear to show significant polarization in either the $R$-band or the $I$-band or both; they are presented in Table~\ref{polarimetry3} and are clearly marked in all the Figures of this paper. Along with their spectral types, polarization degrees, position angles of the plane of vibration, and their associated uncertainties, Table~\ref{polarimetry3} also lists the quantity $P/\sigma_P$, the equivalent widths (if detected spectroscopically) of H$\alpha$ emission and Li\,{\sc i} $\lambda$670.8\,nm absorption, projected rotational velocities and evidence for infrared or near-infrared flux excesses. The spectroscopic information has been gathered from the literature (Ruiz et al$.$ \cite{ruiz97}; Mart\'\i n et al$.$ \cite{martin98}, \cite{martin01b}; Kirkpatrick et al$.$ \cite{kirk99}, \cite{kirk00}; Basri et al$.$ \cite{basri00a}; Schweitzer et al$.$ \cite{schweitzer01}; Liebert et al$.$ \cite{liebert03}; Jayawardhana et al$.$ \cite{jayawardhana03}; Cruz et al$.$ \cite{cruz03}; Bailer-Jones \cite{bailer04}), while the evidence for infrared excesses is based on the recent $L'$ and $M'$ photometric measurements by Liu et al$.$ \cite{liu03} and Golimowski et al$.$ \cite{golimowski04}. We will relate the photometric and spectroscopic properties of each object to its measured polarization in Section~\ref{polarized_dwarfs}.

\begin{figure}
\epsscale{0.75}
\plotone{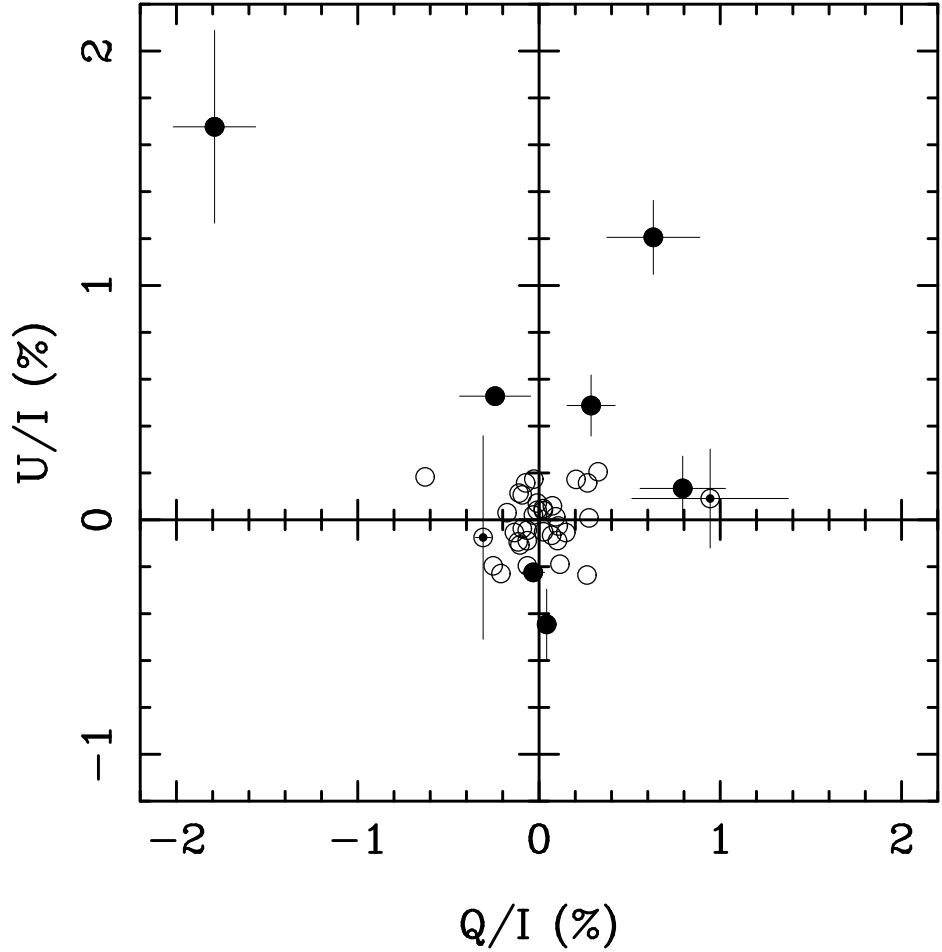}
\caption{The Stokes parameters, $Q/I$ and $U/I$, of our target sample are plotted against each other. All observations are in the Johnson $I$-band centred on 850\,nm and 150\,nm wide. Filled circles stand for the likely polarized dwarfs with detections at $\ge$3\,$\sigma$. The mean values of J2158$-$15 and J2252$-$17 are plotted as circled dots. Error bars of non-detections are avoided for the clarity of the figure.\label{qupol}}
\end{figure}

\begin{figure}
\epsscale{0.75}
\plotone{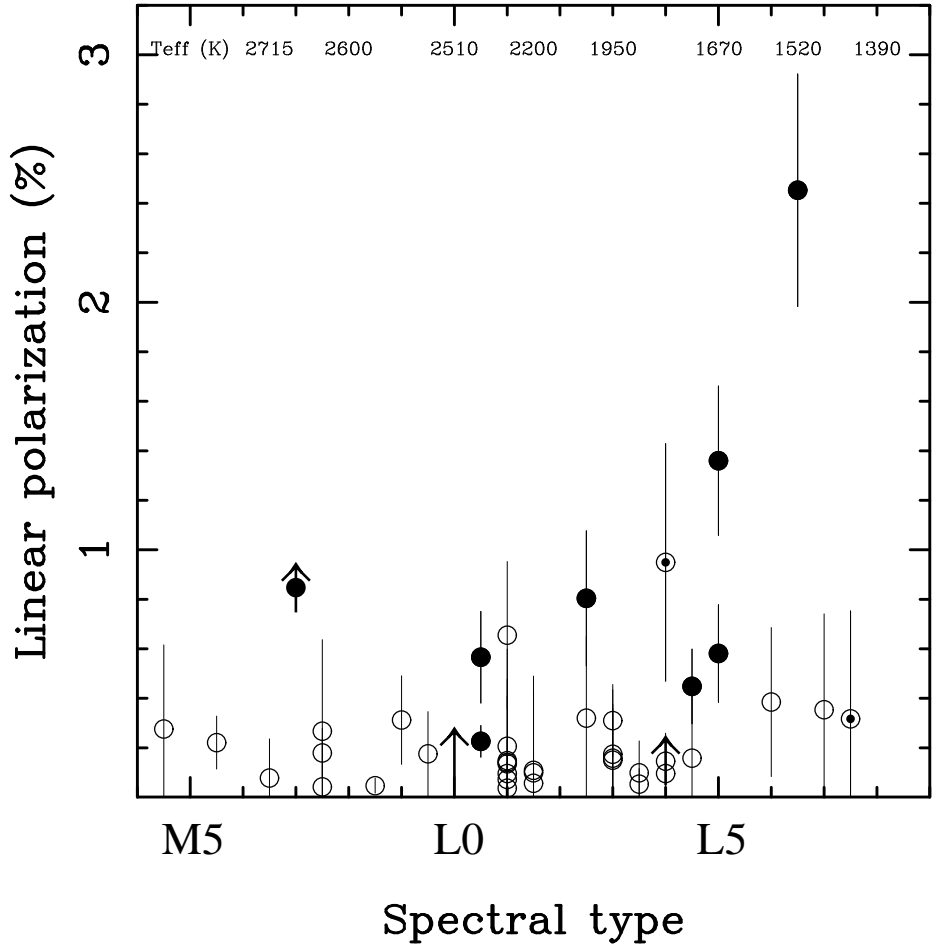}
\caption{$I$-band linear polarization of our target sample against spectral type. Likely polarized sources are plotted as filled circles. Arrows denote lower limits ($P$\,$\ge$\,$|Q/I|$) and open circles stand for non-detections. The mean values of J2158$-$15 and J2252$-$17 are plotted as circled dots. The $T_{\rm eff}$--spectral type calibrations of Dahn et al$.$ \cite{dahn02} (late-M) and Vrba et al$.$ \cite{vrba04} (L types) are given on the top of the figure. The uncertainty in spectral type is half a subclass.\label{ipol}}
\end{figure}

Our $Q/I$ and $U/I$ measurements are plotted in Fig$.$~\ref{qupol}. This plot provides a graphical summary of Tables~\ref{polarimetry1} and \ref{polarimetry2} and illustrates our criterion for polarization. It also shows the symmetrical distribution of the null polarimetric measurements around (0,0), i.e$.$ there is no obvious instrumental bias in our data. The $I$-band polarization degree is depicted against spectral type in Fig$.$~\ref{ipol}. Note that we have labeled the $T_{\rm eff}$--spectral class calibrations of Dahn et al$.$ \cite{dahn02} and Vrba et al$.$ \cite{vrba04} in Fig$.$~\ref{ipol}. Three of our targets (J0036$+$18, J2057$-$02, and J2224$-$01) are in common with the previous work of M\'enard et al$.$ \cite{menard02}. Except for J0036$+$18, which is polarized according to these authors, our values of $I$-band linear polarization are in full agreement with their measurements within 1\,$\sigma$ the error bars. We note, however, that the central wavelength and the passband of the $I_{\rm Bessel}$ filter used by M\'enard et al$.$ is bluer than those of the $I_{\rm Johnson}$ filter used by us. Actually, the passband of the $I_{\rm Bessel}$ filter appears somewhat intermediate between the Johnson $R$- and $I$-bands.

\section{Discussion}

There are several possible mechanisms that could account for the observed linear polarization of ultracool dwarfs: (i) interstellar polarization, (ii) polarization due to strong magnetic fields (Zeeman splitting, synchrotron emission), (iii) scattering by photospheric particulates of dust, and (iv) scattering by the grains of cool shells or disks around the central dwarf. In Sect$.$ \ref{target_selection}, we have argued that an interstellar origin for the linear polarization of our objects is unlikely because of their very short distances. M\'enard et al$.$ \cite{menard02} discussed widely the three former possible origins (they did not mention the fourth possibility), concluding that, pending definitive measurements of the magnetic fields of ultracool dwarfs, scattering by photospheric dust grains remains the most likely origin for the polarization. We also agree with their discussion (and will not repeat it here), coupled with the fact that the presence of circum(sub)stellar disks plays a role in the polarization of the photospheric radiation in very young ultracool dwarfs, as in T\,Tauri stars.

\subsection{Linear polarization and spectral type}

Of the total of 45 targets with spectral types ranging from $\sim$M5V to L7.5V, eleven appear to show some degree of polarization (Table~\ref{polarimetry3}). For these, we have measured $I$-band polarimetric amplitudes in the interval 0.2--2.5\%. These values are in agreement with the predictions of Sengupta \& Krishan \cite{sengupta01}, which were calculated for the case of dust scattering in oblate dwarfs. One of our likely polarized targets is CFHT-BD-Tau\,4, the young brown dwarf of the Taurus star-forming region (Mart\'\i n et al$.$ \cite{martin01b}). As will be discussed below, a photospheric origin for its observed linear polarization is unlikely, and we rule this object out from the following statistical analysis. We will also remove J1610$-$00 because the metallicity of this dwarf is considerably below solar (L\'epine et al$.$ \cite{lepine03}). 

The rate of intrinsically polarized ultracool dwarfs in our sample with spectral types in the interval L0V--L7.5V turns out to be 29\,$\pm$\,9\%, which is below the 50\%~estimated by M\'enard et al$.$ \cite{menard02}. The larger error bars of our measurements have prevented us from easily detecting polarization dregrees smaller than about 0.4\%, which is at least 4 times higher than what M\'enard et al$.$ can measure from their FORS1/VLT data. Observations with better accuracy may confirm polarization of dwarfs that show marginal detection in our study. Hence, the frequency of 29\%~has to be understood as a lower limit on the ocurrence of linear polarization amongst solar metallicity, field L-type dwarfs. This fraction is remarkably high when compared to the frequencies of polarized stars of warmer spectral types, which are typically below 10\%~(see Leroy \cite{leroy99}). Very recently, M\'enard \& Delfosse \cite{menard04} have analized 20 nearby field dwarfs spanning spectral classes M1 to M6, and have found that all their measurements are compatible with a null polarization. Furthermore, in our sample there are 9 field dwarfs with types ranging from mid-M to very late-M (M9.5V). If the $I$-band polarization rate of 29\%~were valid for M dwarfs, we would expect to find 2--3 M-type polarized stars in our sample. However, we do not detect significant polarization in any of the M-type field targets, which suggests that, at 95\%~confidence, less than 28\%~of M dwarfs are polarized. Hence, at far-red optical wavelengths, the frequency of highly polarized L-type dwarfs is clearly larger than that of M dwarfs. 

This is an evidence for the existence of efficient dust scattering processes in L-type atmospheres. As will be mentioned in Section~\ref{activity}, L dwarfs are on average less active than M stars (Gizis et al$.$ \cite{gizis00}). In addition, L-type atmospheres are more neutral and present higher resistivities because of their lower temperatures; hence, we do not expect the magnetic fields of L dwarfs to be stronger than those of M stars (see Mohanty et al$.$ \cite{mohanty02}). Furthermore, the weakening of oxides (TiO, VO) in late-M and early-L types and of hydrides (FeH, CrH) at cooler types indicates that the metals like Ti, V and Fe gradually vanish from the gas phase (Mart\'\i n et al$.$ \cite{martin99a}; Kirkpatrick et al$.$ \cite{kirk99}). This occurs at temperatures for which models predict the formation of condensates (Tsuji et al$.$ \cite{tsuji96}; Allard et al$.$ \cite{allard01}). These phenomena coupled with our polarimetric detections and those of M\'enard et al$.$ \cite{menard02} support the presence of significant amounts of dust in ultracool atmospheres.

Also relevant is the study of the degree of linear polarization as a function of the L subtypes (or $T_{\rm eff}$ between 2500 and 1400\,K). Figure~\ref{ipol} depicts our results. We shall now focus on the dwarfs with detected polarization according to our criterion. There is a hint for cooler dwarfs displaying larger linear polarization degrees, i.e$.$ the polarization of the $I$-band radiation seems to be more powerful at low temperatures. The dwarfs J2244$+$20 (L6.5V) and J1507$-$16 (L5V), two of the latest objects in our sample, show the largest measured linear polarizations, $P$\,=\,2.5\,$\pm$\,0.5\%~and 1.36\,$\pm$\,0.30\%, respectively. These values contrast with the moderate polarization observed in the much warmer dwarf J1707$+$43 (L0.5V, $P$\,=\,0.23\,$\pm$\,0.06\%). Intermediate objects display polarizations between those of the L0.5V and L6.5V dwarfs. Such a trend was also mentioned by M\'enard et al$.$ \cite{menard02}. Either different sources or mechanisms are inducing the polarization or the trend is possibly related to the rate of grain formation and the vertical distribution of the clouds. As pointed out by Sengupta \cite{sengupta03}, there are several possibilities to polarize the radiation via scattering (spherical and nonspherical grains in spherical and oblate photospheres, random distribution of condensates, and presence of dust bands). The more possibilities working at the same time, the larger the polarization. The trend might also be associated to a faster rotation of the coolest dwarfs. However, Mohanty \& Basri \cite{mohanty03} and Bailer-Jones \cite{bailer04} noted that the projected rotational velocities of L0--L8 field dwarfs show no obvious trend with spectral type. Nevertheless, more polarimetric observations and higher accuracies are required to confirm this dependency. Theoretically, the metal content is also a key parameter in the rate of grain formation (Allard et al$.$ \cite{allard01}); hence, changes in the metallicity of ultracool dwarfs may also account for possible scatter in the polarization amplitudes of Fig$.$~\ref{ipol}. 

From our data, we can also investigate the frequency of $I$-band polarized dwarfs per intervals of spectral type. Three intervals have been chosen: from M4.5V to M9.5V, from L0V to L3V, which corresponds to $T_{\rm eff}$\,$\sim$\,2500--1900\,K, and from L3.5V to L8V, corresponding to $T_{\rm eff}$\,$\sim$\,1900--1300\,K. Table~\ref{freq} summarizes our results (note that CFHT-BD-Tau\,4 and J1610$-$00 are excluded from the statistics). For the early L-types, there are 3 positive detections out of 20 investigated dwarfs (15\,$\pm$\,9\%), while the rate of linear polarization increases notably for the late types (6 detections out of 14 dwarfs, i.e$.$ 43\,$\pm$\,17\%). Similar statistics are obtained from the results of M\'enard et al$.$ \cite{menard02}, although the number of studied objects in their work is relatively small. We note that in our analysis, exposure times were set for each target to approximately compensate for different brightness. Hence, the minimum detection level of polarization and the mean polarimetric errors of both spectral intervals are alike. Despite this, the incidence of high $I$-band linear polarization appears to be a factor of 2--3 larger in the coolest L spectral types than in the early L classes. However, we caution that this result may be biased by the uncertainty of the measurements and the fact that very cool atmospheres may be more polarized: to detect low to moderate degrees of polarization, extremely good signal-to-noise ratios are needed. This is, for example, the case of J1707$+$43, for which we have measured a polarization degree well below 0.4\%.

We also caution that there are other biases in our sample possibly influencing our statistical analysis. Young objects are significantly brighter and easier to detect. This also applies to unresolved multiple systems, particularly those comprised of nearly equal mass components. We have selected our targets to be amongst the brightest sources for their spectral classes. The tidal effects of very close binaries suggest nonspherical shapes, which contributes to increase the net polarization. The binary frequency of field ultracool dwarfs is recently determined to be about 15\%~in the separation range 1.6--16\,AU (Gizis et al$.$ \cite{gizis03}). Very little is known for closer orbits, although Gizis et al$.$ \cite{gizis03} have suggested that the binary fraction is $\sim$5\%~for separations less than 1.6\,AU. We note that we have resolved one double L dwarf in our sample, J1705$-$05 (L4), for which we do not detect $I$-band polarization ($P$\,$\le$\,0.2\%). The separation is estimated at less than 1.3\,arcsec (i.e$.$ less than about 14\,AU at the distance of 10.7\,pc, Kendall et al$.$ \cite{kendall04}). Further data (e.g$.$ proper motions) are needed to confirm the physical link of J1705$-$05. On the other hand, some of the L dwarfs in our sample might have young ages. As very low-mass stars do at early stages of evolution, very young brown dwarfs also happen to harbor disks from which they can accrete (e.g$.$ Natta \& Testi \cite{natta01}; Jayawardhana et al$.$ \cite{jayawardhana02}). Disks efficiently polarize the photons from the central object. We cannot provide precise age estimates for each object in our sample because many lack astrometric parallaxes. However, warm disks around low-mass stars and brown dwarfs are typically observed at ages below 10\,Myr (e.g$.$ Brandner et al$.$ \cite{brandner00}; Barrado y Navascu\'es \& Mart\'\i n \cite{barrado03}). The frequency of such young objects among K- and M-type stars in the solar neighborhood ($d$\,$\le$\,50\,pc) is very low ($\sim$10\%). Hence, we do not expect that the biases due to binarity and extreme youth are critical in our analysis. We will address these topics again in Section~\ref{polarized_dwarfs}, where we discuss each polarized dwarf separately. 

In general, and on the assumption that our targets have similar ages (with very few exceptions), our polarimetric observations of late-M and L-type dwarfs are qualitatively in agreement with the theoretical assumption that a plethora of dust grain formation and condensation takes place in the outer atmospheric layers of objects with spectral type later than M7 (Tsuji et al$.$ \cite{tsuji96}). As clouds form in progressively cooler objects, they become optically thicker and form deeper within the atmosphere. In addition, as $T_{\rm eff}$ decreases, more grain species are formed. Hence, dust and condensates must play a role in the output energy distributions (absorption, scattering, polarization, chemistry, thermal structure). The tropospheric weather pattern predicted for brown dwarfs (Schubert \& Zhang \cite{schubert00}) can more easily produce inhomogeneities in the distribution of the clouds by creating local clearings since the turbulent motions are greater. This would make polarization more likely. On the other hand, optical and near-infrared spectra of the coolest dwarfs ($T_{\rm eff}$\,$<$\,1300\,K, the T domain) indicate that, for such low temperatures, dust remains in the form of a thin cloud very deep in the photosphere, i.e$.$ dust grains are segregated and precipitated (e.g$.$ Allard et al$.$ \cite{allard01}; Burrows et al$.$ \cite{burrows02}; Tsuji et al$.$ \cite{tsuji04}). In this way, dust particles are neither affecting the atmospheric thermal structure nor blocking nor polarizing the emergent radiation. There are no T-dwarfs in our sample. Furthermore, the coolest objects in our study lie close to the L--T transition, for which models predict the largest dusty coverage of the photosphere. It would be stimulating to extend the polarimetric studies toward cooler types. There is evidence in early T dwarfs (T0--T3) of effects of clouds on the emergent spectra (see Marley et al$.$ \cite{marley02}; Burrows et al$.$ \cite{burrows02}; Burgasser et al$.$ \cite{burgasser02b}; Tsuji et al$.$ \cite{tsuji04}). No polarization from scattering is expected for later types.

\subsection{Linear polarization and metallicity}

We note that the atmospheric chemical abundance, which may play a critical role in polarization, is not well determined for any of the objects in our sample. 

Nevertheless, we have included in our study a suspected low-metallicity L-type dwarf, J1610$-$00 (L\'epine et al$.$ \cite{lepine03}), which was found in a proper motion survey. The near-infrared colors of this object are bluer than expected for solar metallicity L dwarfs (Zapatero Osorio et al$.$ \cite{osorio04}). This photometric property is consistent with the predictions of theoretical models for metal-depleted ultracool objects. The low-resolution spectrum shown in L\'epine et al$.$ \cite{lepine03} also indicates that J1610$-$00 has a low metal content. In our sample, there is another dwarf, J1721$+$33 (L3V), displaying bluer near-infrared colors than average by about 0.2\,mag (Cruz et al$.$ \cite{cruz03}). In addition, J1721$+$33 has significant proper motion, suggesting that it is part of a low-metallicity population. Our polarimetric measurements for this object and J1610$-$00 are consistent with zero polarization. The low number of suspected metal-depleted ultracool dwarfs in our sample does not allow us to discuss the dependence of polarization on metallicity. However, it is expected that the photons of metal-depleted dwarfs are less polarized than those of metal-rich dwarfs because grain formation is far reduced under low metallicity conditions (Allard et al$.$ \cite{allard01}).

\subsection{Linear polarization and photometric variability, activity and rotation \label{activity}}

\begin{figure}
\epsscale{0.75}
\plotone{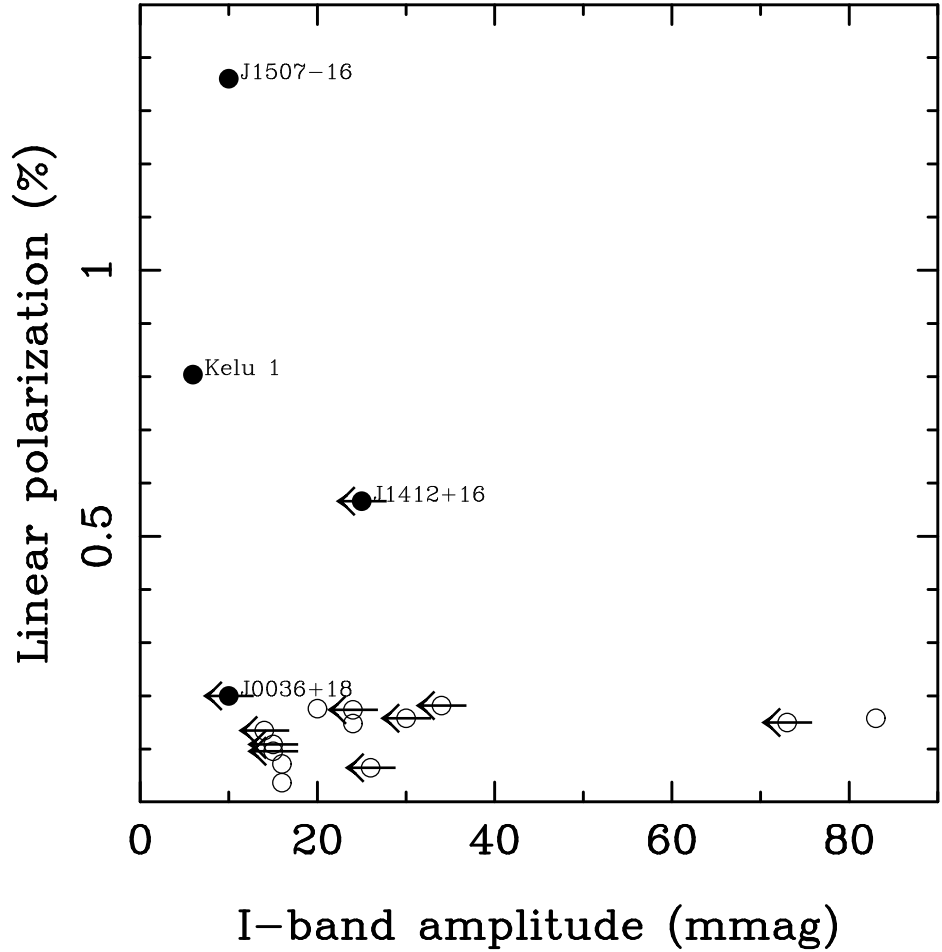}
\caption{$I$-band linear polarization against the amplitude of the photometric variability at the same wavelengths. Polarimetric data of our work and of M\'enard et al$.$ \cite{menard02} are plotted in the diagram. Error bars have been avoided for the sake of clarity. Four polarized dwarfs are labeled and plotted as filled circles. Polarimetric non-detections are shown with open circles. Upper limits on the $I$-band photometric variability (as reported in the literature) are indicated with left arrows. Three likely polarized dwarfs display rather small photometric amplitudes ($\le$10\,mmag).\label{ivar}}
\end{figure}

A total of 18 dwarfs in our sample have been photometrically monitored by other groups to investigate $I$-, $J$- or $K$-band variability. Table~\ref{photvar} presents a summary. The fourth column of the Table indicates whether broad-band photometric variability has been detected (the filter is also indicated), and the last column provides the bibliographic references. Seven dwarfs have strong variability detections, four display weak variability, and a few show periodic modulations. We note that the photometric data gathered from the literature are not simultaneous to our polarimetry. The amplitudes of the $I$-band light curves obtained from the literature and from Goldman (2004, priv$.$ communication) are depicted as a function of linear polarization in Fig$.$~\ref{ivar}. Detections and non-detections are plotted with different symbols. Weak photometric variabilities are considered non-detections in the figure and in the following discussion.

Whether the formation, distribution and evolution of photospheric dust clouds (``meteorology'') is causing the reported photometric variability is not unequivocally confirmed. Yet, as discussed by the various authors, there are reasons to believe in a connection between photometric variability and dust clouds (see discussions by Mart\'\i n et al$.$ \cite{martin01a}; Bailer-Jones \cite{bailer02}; Bailer-Jones \& Lamm \cite{bailer03}). We would also expect some correlation between polarization and variability in ultracool atmospheres. Four likely polarized dwarfs have been photometrically monitored (see Table~\ref{photvar}). They are labeled in Fig$.$~\ref{ivar}. Two of them (Kelu\,1 and J1507$-$16) are reported to be variable, and the other two (J0036$+$18 and J1412$+$16) have upper limits on the amplitude of their photometric variations. It is worth mentioning that Kelu\,1, J1507$-$16 and J0036$+$18 show the smallest $I$-band amplitudes ($\le$10\,mmag) in the sample, which contrasts with their high linear polarizations. On the other hand, the five dwarfs with larger photometric amplitudes ($\ge$10\,mmag) and confirmed variability (open circles without arrows in Fig$.$~\ref{ivar}) do not display significant polarization ($P$\,$\le$\,0.2\%). S$.$ Sengupta (2004, priv$.$ communication) has suggested that flux variability, if due to dust activity, needs sufficiently optically thick dust clouds. In this medium, polarization would arise by means of multiple scattering processes, which in turn would reduce the degree of polarization as compared to single scattering of photons in optically thin clouds (see Sengupta \cite{sengupta03}). Thus, dwarfs with strong photometric variability may have less polarization because of multiple scattering. On the other hand, dwarfs with weak or no photometric variability (static dust clouds or optically thin cloud layer) may give rise to higher polarization. Nevertheless, we strongly remark that more data are needed before we can confirm (or discard) any relation between polarization and broad-band photometric variability. At best, polarimetric observations and photometric monitoring should be carried out simultaneously. Furthermore, because ultracool dwarfs rotate very rapidly and because of fast tropospheric motions (Schubert \& Zhang \cite{schubert00}), the patterns of clouds are expected to change in a few rotational periods, producing modifications in the amplitudes and directions of the polarization. Hence, the detection of variations in both polarimetry and photometric monitoring will provide a strong case for the evolution of dust clouds (``weather'') in ultracool atmospheres.

So far it is not possible to draw any conclusions about the relation between optical polarization and activity as measured from H$\alpha$ emission, flares, X-ray or radio emission. There are very few data on this respect in the literature, and what is found appears to be contradictory. Berger \cite{berger02} has detected radio flares and radio constant emission from BRI\,0021 and J0036$+$18 in our sample, suggesting that radio activity is present between M-types and L3.5V. This contrasts with the observed drop in persistent H$\alpha$ activity beyond spectral type M7 (Burgasser et al$.$ \cite{burgasser00}). Moreover, neither BRI\,0021 nor J0036$+$18 show H$\alpha$ emission (less than 0.5\,\AA~pseudoequivalent width), except for very rare occasions (a flare in BRI\,0021 was detected once, Reid et al$.$ \cite{reid99}). From our data and according to M\'enard et al$.$ \cite{menard02}, J0036$+$18 exhibits some polarization while BRI\,0021 does not. The high polarization measured in J1507$-$16 does not correlate with the non-detection of radio emission (Berger \cite{berger02}). When compared to warmer stars, ultracool dwarfs (except for the very young ones) appear to be rather inactive in X-rays despite of their rapid rotation and, in some cases, moderately strong activity at radio wavelengths (Mart\'\i n \& Bouy \cite{martin02}). The source responsible for polarization does not seem to be strongly related to magnetic fields, since the intensity of these fields as inferred from radio and X-ray activity is quite low and the atmospheres of ultracool dwarfs are rather neutral. 

\begin{figure}
\epsscale{0.75}
\plotone{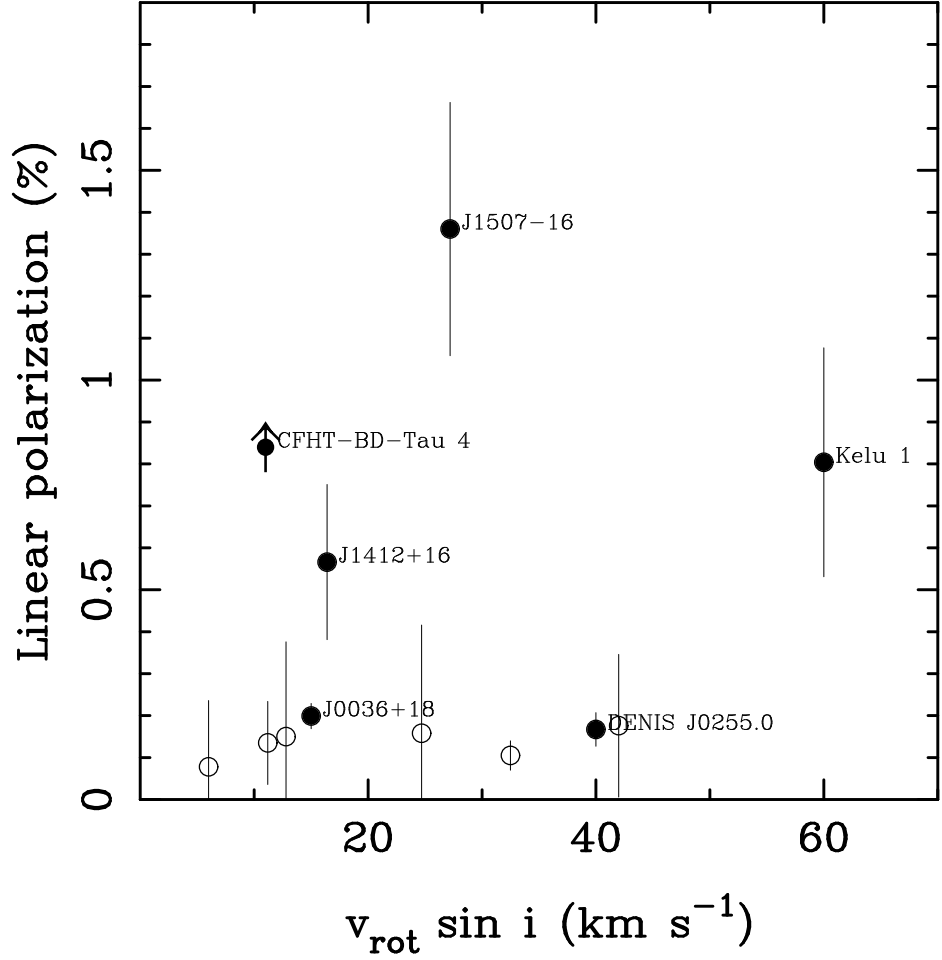}
\caption{$I$-band linear polarization against projected rotational velocity of field dwarfs. Likely polarized dwarfs from our work and from M\'enard et al$.$ \cite{menard02} are labeled and plotted as filled circles. Polarimetric non-detections and upper limits are shown with open circles and arrows, respectively. The apparent lack of correlation between the degree of linear polarization and the projected rotational velocity may be due to the uncertainty introduced by the unknown rotation angle and the sparse number of data. \label{veloc}}
\end{figure}

Figure~\ref{veloc} displays our polarimetric measurements as a function of projected rotational velocities ($v$\,sin\,$i$), which have been taken from the literature (see references above, and Basri \& Marcy \cite{basri95}). The great majority of the ultracool dwarfs display rapid rotation (typically $v$\,sin\,$i$\,$\sim$\,10--60\,km\,s$^{-1}$), despite the fact they do not seem to be very active objects (Basri \cite{basri00b}). Fast rotational velocities would impose deviations from sphericity in the shape of the objects (i.e$.$ asymmetry), favoring the detection of polarization in dusty atmospheres (Sengupta \& Krishan \cite{sengupta01}). We fail to observe any correlation in Fig$.$~\ref{veloc}. However, this may be due to the uncertainty introduced by the unknown equatorial inclination of $v$\,sin\,$i$, and the relatively poor statistics, i.e$.$ there are rather few objects for which polarimetric observations and velocities are available. 

\subsection{The likely and possible polarized dwarfs \label{polarized_dwarfs}}

Finally, we will discuss each of the polarized dwarfs separately. The astrometric parallaxes have been obtained for nine of our target dwarfs (Dahn et al$.$ \cite{dahn02}; Vrba et al$.$ \cite{vrba04}), allowing us to produce an HR diagram as the one depicted in Fig$.$~\ref{hr}. To incorporate CFHT-BD-Tau\,4 into the Figure, we have converted its apparent $K$-band magnitude into the absolute value using the cannonical distance to the Taurus star-forming region (140\,pc). Three isochrones of 10, 100 and 1000\,Myr from Chabrier \& Baraffe \cite{chabrier00} are overplotted onto the data. The mass intervals covered by the Figure are as follows: 7--40\,$M_{\rm Jup}$ (Jovian masses, 10\,Myr), 20--70\,$M_{\rm Jup}$ (100\,Myr), and 50--90\,$M_{\rm Jup}$ (1\,Gyr). These state-of-the-art evolutionary models provide magnitudes in the filters of interest. To transform predicted $T_{\rm eff}$'s into spectral types, the calibrations given by Dahn et al$.$ \cite{dahn02} for late-M types and Vrba et al$.$ \cite{vrba04} for the L classes have been applied. 

\subsubsection{2MASS\,J00361617$+$1821104}

According to M\'enard et al$.$ \cite{menard02}, this L3.5V dwarf ($T_{\rm eff}$ $\sim$ 1900\,K) shows some polarization ($P$\,=\,0.20\,$\pm$\,0.03\%) in the $I_{\rm Bessel}$-band, which is centred on 768\,nm. On the contrary, our $I_{\rm Johnson}$ data suggest that there is very little or no polarization at redder wavelengths (850\,nm). At the same time, we do detect significant polarization at short wavelengths, in the $R$-band (641\,nm). This is worthy of especial mention. The dust properties (e.g$.$ refractive index and scattering cross-section) depend on wavelength, i.e$.$ they vary from one wavelength to another. Additionally, there is a hint for the variability of the $R$-band polarization angle, as inferred from Table~\ref{polarimetry3}. Our data indicate that the degree of linear polarization is possibly a few times larger at around 641\,nm than at about 850\,nm. This and the measurement of M\'enard et al$.$ \cite{menard02} suggest that the polarization observed in J0036$+$18 systematically decreases with increasing wavelength, providing evidence for the presence of dust grains in the photosphere of this L-type dwarf since this behavior may not be accounted for by other mechanisms, e.g$.$ magnetic fields. This polarimetric property may also shed new light on the size of the particles responsible for the observed polarization. 

As discussed by M\'enard et al$.$ \cite{menard02}, the intensity of the magnetic field that might be present in J0036$+$18 as inferred from its radio detection (135\,$\mu$Jy at 8.46\,GHz, Berger \cite{berger02}), is not powerful enough to induce a detectable linear polarization of the emergent optical radiation. In addition, J0036$+$18 was not detected as variable (upper limits of 9, 16 and 25\,mmag on the $I$, $J$ and $K_s$ variability, respectively) by the photometric monitoring programs of Gelino et al$.$ \cite{gelino02} and Caballero et al$.$ \cite{caballero03}. On the other hand, J0036$+$18 does not possess any infrared flux excesses at 3.8\,$\mu$m or 4.7\,$\mu$m (Golimowski et al$.$ \cite{golimowski04}), suggesting that there is no massive nor warm disk around this dwarf. Moreover, the optical spectrum of J0036$+$18 does not show the Li\,{\sc i} feature at 670.8\,nm, which indicates that the age of the object is older than several hundred Myr (Chabrier \& Baraffe \cite{chabrier00}). This is also consistent with the location of this dwarf in the HR diagram of Fig$.$~\ref{hr}. The disk is very likely dissipated at these ages and does not contribute significantly to the polarization. Hence, the most likely origin of the measured polarization in J0036$+$18 is related to the formation of dust clouds within the object's atmosphere. From theoretical considerations, the degree of polarization due to single and multiple scattering will be more in the blue wavelengths if the grain size is very small (Sengupta \& Krishan \cite{sengupta01}). By inspecting Fig.~1 of Sengupta \cite{sengupta03}, the models that better reproduce the trend of the polarimetric observations of J0036$+$18 are those computed for atmospheres with dust particle sizes less than 1\,$\mu$m in diameter. The synthetic spectra provided by many groups (e.g$.$ Tsuji et al$.$ \cite{tsuji96}, \cite{tsuji04}; Allard et al$.$ \cite{allard01}) are obtained assuming size distributions of the grains between $\sim$0.006\,$\mu$m and 0.25\,$\mu$m (submicron range), which is consistent with our result. However, the recent calculations of Woitke \& Helling \cite{woitke04} predict grain growths up to 30 and 400\,$\mu$m at the deepest cloud layers. It would be desiderable to obtain more polarimetric data at longer and shorter wavelengths to provide a deeper study of the physical properties of the dust grains. If the particles are very small, no polarization is expected beyond 1.3\,$\mu$m (i.e$.$ the $J$-band). 

\subsubsection{2MASS\,J01410321$+$1804502}

Recently confirmed as a L4.5V dwarf by Wilson et al$.$ \cite{wilson03}, as far as we know there is no more information on this object available in the literature. More data are required to confirm the $I$-band polarimetric detection. The three measurements shown in Table~\ref{polarimetry1} are consistent with each other within 1\,$\sigma$ the uncertainties, suggesting little variability in scales of days.

\subsubsection{2MASS\,J01443536$-$0716142}

Liebert et al$.$ \cite{liebert03} report the detection of a flare event in this L5V-type dwarf. This object displayed strong H$\alpha$ emission which rapidly declined in about 15 minutes. Based on their spectroscopic data, the authors concluded that L-dwarfs are observed in strong flares only occasionally. Further polarimetric observations are needed to confirm our (marginal) detection. The measurements obtained on four different epochs indicate little variability within 2 $\sigma$ the uncertainties. 

\subsubsection{CFHT-BD-Tau\,4} 

This is one of the warmest dwarfs in our sample (spectral class M7), and is also the youngest object. It is suspected to be a member of the Taurus star-forming region, with an age estimated at less than a few Myr. Its location in Fig$.$~\ref{hr} clearly reveals that CFHT-BD-Tau\,4 is overluminous as compared to the field sequence. Previous photometric and spectroscopic studies identify this object as a very young brown dwarf, which is probably obscured by the presence of a surrounding shell from which the central object is intensively accreting. CFHT-BD-Tau\,4 shows moderate extinction ($A_V$\,$\sim$\,3\,mag), very strong and variable H$\alpha$ emission (Mart\'\i n et al$.$ \cite{martin01b}; Jayawardhana et al$.$ \cite{jayawardhana03}), X-ray activity (Mokler \& Stelzer \cite{mokler02}) and infrared excesses (Liu et al$.$ \cite{liu03}; Pascucci et al$.$ \cite{pascucci03}), as do ordinary T\,Tauri stars. Gorlova et al$.$ \cite{gorlova03} determined a rather low surface gravity for this object, which is consistent with very young ages. Additionally, the recent detection of millimeter dust emission (Klein et al$.$ \cite{klein03}) and the multiwavelength study of Pascucci et al$.$ \cite{pascucci03} support the presence of circum(sub)stellar dust of about a few Earth masses in the form of a disk (the total mass of the disk is estimated at about 1 Jupiter mass, if we extrapolate the dust masses to disk masses assuming a gas-to-dust ratio of 100). We detect significant linear polarization in both $R$- and $I$-band wavelengths, which can be attributed to the disk around the central object, supporting the T\,Tauri scenario. We remark that other (older) late-M dwarfs in our sample do not show any evidence for linear polarization. Our data suggest that the degree of polarization of CFHT-BD-Tau\,4 shows a trend in the sense that the amount of polarized radiation increases with wavelength. This trend is also observed in various T\,Tauri stars (e.g$.$ V\,410\,Tau, Mekkaden \cite{mekkaden99}). This behavior is different from that of J0036$+$18, being a hint toward large grain sizes in the disk of CFHT-BD-Tau\,4. This result is in agreement with the very recent infrared observations of Apai et al$.$ \cite{apai04}. These authors determine that the silicate feature of CFHT-BD-Tau\,4 is dominated by emission from 2\,$\mu$m amorphous olivine grains. This brown dwarf provides compelling evidence that young substellar objects undergo a T\,Tauri-like accretion phase similar to that in low-mass stars. We do not detect photopolarimetric variability within 1\,$\sigma$ the error bars on scales of up to three days. However, it would be interesting to monitor this object (simultaneous photometry and polarimetry) in order to constrain the geometry of the circum(sub)stellar material. 

\subsubsection{Kelu\,1} 

\begin{figure}
\epsscale{0.75}
\plotone{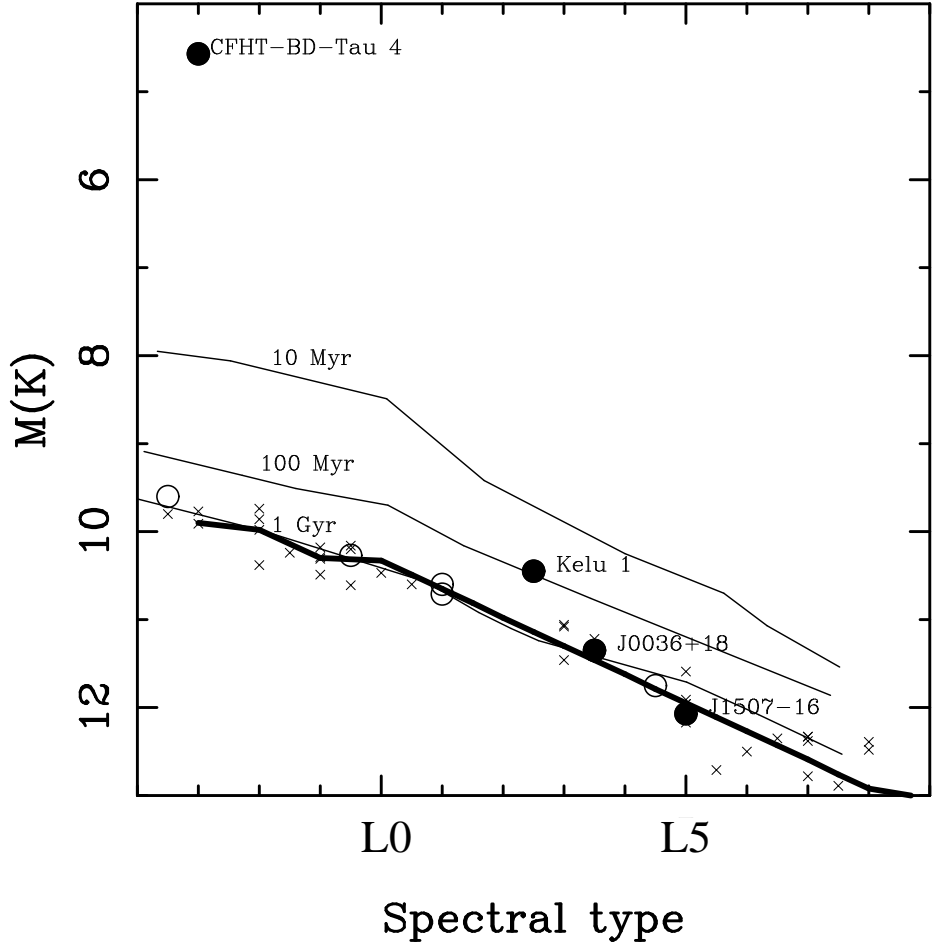}
\caption{$K$-band absolute magnitude against spectral type of sources with trigonometric parallaxes. Likely polarized dwarfs are labeled (filled circles) and unpolarized objects are plotted as open circles. The error bar of the photometry is of the order of the size of these symbols. The average location of the field sequence is plotted as a thick line (Dahn et al$.$ \cite{dahn02}; Vrba et al$.$ \cite{vrba04}); crosses denote individual field dwarfs. Three isochrones (10\,Myr, 100\,Mry and 1\,Gyr) from the Lyon models (Chabrier \& Baraffe \cite{chabrier00}) are also plotted as thin lines. CFHT-BD-Tau\,4 appears quite overluminous due to its very young age (a few Myr or less). The other outlier in this diagram is Kelu\,1, which lies close to the 100\,Myr isochrone. It may be a binary of similar components (not likely, see text) or a relatively young brown dwarf (a few hundred Myr) of about 35\,$M_{\rm Jup}$.\label{hr}}
\end{figure}

To our knowledge, this L2.5V-type brown dwarf ($T_{\rm eff}$\,$\sim$\,2030\,K) shows the largest projected rotational velocity amongst all known ultracool dwarfs (Basri et al$.$ \cite{basri00a}), indicating that it is indeed a very fast rotator. Hence, it is expected that Kelu\,1 shows an oblate shape, which coupled with the formation of dust clouds favor the detection of polarization. Kelu\,1 has also been found to be photometrically variable with a periodicity of about 1.8\,h (Clarke et al$.$ \cite{clarke02a}); nevertheless, the amplitude of the $I$-band light curve is small (roughly 6\,mmag). Further spectrocopic follow-up of this object reveals a modulated variability of the H$\alpha$ line intensity; the line is always detected in rather low emission (Clarke et al$.$ \cite{clarke03}). Kelu\,1 appears overluminous in the diagram of Fig$.$~\ref{hr}, which suggests that this object is either a binary of similar components or a young object or both. However, high-resolution direct imaging and radial velocity searches for companions have failed to resolve Kelu\,1 (Mart\'\i n et al$.$ \cite{martin99b}; Clarke et al$.$ \cite{clarke03}), imposing upper limits on the separation of the binary ($\le$0.3\,arcsec) and on the mass of the companion ($M$sin$i$\,$\le$\,10\,$M_{\rm Jup}$). Youth is quite likely because the Li\,{\sc i} resonance doublet is detected in the optical spectrum, which implies young ages and consequently relatively lower gravities than field dwarfs of similar types. Nevertheless, Kelu\,1 is not as young as CFHT-BD-Tau\,4 since no infrared excesses are observed in the former brown dwarf (Golimowski et al$.$ \cite{golimowski04}), it is not located in any star-forming region, and the alkali lines of Kelu\,1's optical spectrum appear relatively strong (McGovern et al$.$ \cite{mcgovern04}). Using evolutionary models as those illustrated in Fig$.$~\ref{hr}, the age of Kelu\,1 is estimated at a few hundred Myr, and its mass is in the interval 30--50\,$M_{\rm Jup}$ (35\,$M_{\rm Jup}$ at the age of 100\,Myr). As pointed out by Sengupta \cite{sengupta03}, for a given rotational velocity, the degree of polarization is higher for objects with comparatively lower surface gravity. Interestingly, various groups of authors provide different pseudo-equivalent width measurements of the Li\,{\sc i} $\lambda$670.8\,nm feature (from slightly less than 1\,\AA~up to 4.7\,\AA, see Mart\'\i n et al$.$ \cite{martin98}, and references therein). This may be indicative of some sort of activity that changes the strength of the TiO molecular absorptions (e.g$.$ different coverage of dust clouds). It is also possible that Kelu\,1 is burning lithium if its age is around a few hundred Myr. From our data, Kelu\,1 shows a detectable polarization in the $I$-band. The fast rotation, the formation of clouds of condensates, and the relatively atmospheric low gravity may account for the measured polarization amplitude.

\subsubsection{2MASS\,J14122449$+$1633115} 

Our polarimetric data suggest that this L0.5V-type dwarf shows some polarization. However, the detection is marginal, with $P/\sigma_P$\,=\,3.0. Further data would be needed to confirm or discard our result. Gelino et al$.$ \cite{gelino02} did not find any evidence for photometric variability in J1412$+$16, imposing an upper limit on the $I$-band variability of 25\,mmag. In addition, this object shows moderate H$\alpha$ emission (Kirkpatrick et al$.$ \cite{kirk00}).

\subsubsection{2MASS\,J15074769$-$1627386} 

This L5V-type dwarf shows one of the highest degrees of $I$-band linear polarization in our sample ($P/\sigma_P$\,=\,4.5). Because of the proximity of this object (parallactic distance of 7.3\,pc, Vrba et al$.$ \cite{vrba04}), the interstellar medium is not likely contributing to the observed polarization. From optical spectroscopic observations, J1507$-$16 does not show H$\alpha$ emission, and has efficiently depleted its original lithium content. This suggests, as for J0036$+$18, that the age of J1507$-$16 is older than several hundred Myr. From its location in Fig$.$~\ref{hr}, the most likely age of this object is very similar to that of the great majority of field dwarfs with similar spectral types ($\sim$1--5\,Gyr). The $L'$ and $M'$ photometric data of Golimowski et al$.$ \cite{golimowski04} indicate that there are no infrared excesses associated to this object. And Berger \cite{berger02} reported an upper limit on the radio emission of 58\,$\mu$Jy at 8.46\,GHz, suggesting that if any magnetic field exists, it has to be rather weak, unable to polarize the photospheric optical photons. Thus, the measured polarization is very likely intrinsic to the physical properties of the ultracool atmosphere of J1507$-$16. According to Bailer-Jones \cite{bailer04}, the projected rotational velocity of this object is rather high, which would impose a deviation from sphericity in its shape. A significant oblateness and a heterogeneous distribution of dust clouds on the surface might account for the relatively large degree of polarization. It is worth noting that while J1507$-$16 did not show convincing photometric variability in the $I$-band monitoring program of Koen \cite{koen03}, the CLOUDS collaboration (Goldman \cite{goldman03}) has detected a 1\%~variation ($I$-band) over a few hours (Goldman 2004, priv$.$ communication). More polarimetric data are needed to study changes in the amplitude and direction of the polarization, which in combination with the $I$-band photometric light curves, would confirm the presence of dust clouds in the photosphere.

\subsubsection{2MASS\,J17073334$+$4301304} 

It shows strong H$\alpha$ emission in the spectrum of Cruz et al$.$ \cite{cruz03}. As pointed out by these authors, this L0.5V dwarf should be monitored to check whether it was observed during a period of unusually high activity. We note that the Li\,{\sc i} absorption line at $\lambda$670.8\,nm is not apparent in Fig$.$~8 of Cruz et al$.$ \cite{cruz03}, indicating that the age of this object is larger than several hundred Myr. Our two measurements taken on different epochs do not suggest polarimetric variability.

\subsubsection{2MASS\,J21580457$-$1540098} 

Despite the fact that the averaged polarimetric value shown in Table~\ref{polarimetry2} does not comply with our criterion of polarization, we consider this L4V dwarf to be a likely polarized source. The two separated measurements listed in Table~\ref{polarimetry1} are indicative of the presence of polarized photons. However, these measurements differ by more than 3 $\sigma$ the uncertainties, suggesting some kind of polarimetric variability. The baseline of the two epochs of observations is 0.96\,yr. To the best of our knowledge, this dwarf has not been photometrically monitored by any group of observers. We note that the averaged polarimetric value is plotted in Figs$.$~\ref{qupol} and~\ref{ipol}, and that only the most likely detection (largest $P/\sigma_P$) is listed in Table~\ref{polarimetry3}. The pseudo-equivalent widths of the atomic lines observed in low-resolution near-infrared spectra are very similar to those of field dwarfs (McLean et al$.$ \cite{mclean03}), suggesting that J2158$-$15 has the typical age of other field objects of related spectral classes. 

\subsubsection{2MASS\,J22443167$+$2043433} 

This L6.5V dwarf is a quite peculiar object. It displays the highest $I$-band linear polarization degree in our sample ($P$\,=\,2.5\,$\pm$\,0.5\%), for which we have determined a relatively high $P/\sigma_P$. As pointed out in the literature, J2244$-$01 is nearly 0.3--0.5\,mag redder in optical, near-infrared and infrared colors than all other L dwars (Dahn et al$.$ \cite{dahn02}; Golimowski et al$.$ \cite{golimowski04}; Knapp et al$.$ \cite{knapp04}). In addition, McLean et al$.$ \cite{mclean03} have found that this object has unusual near-infrared spectral features (very weak atomic lines and the peak flux in the $J$-band is less than at $H$ and $K$). In contrast, the optical low-resolution spectrum of J2244$-$01 depicted in Fig$.$~25 of McLean et al$.$ \cite{mclean03} appears similar to other dwarfs of related spectral types. The weak atomic lines observed in the near-infrared spectrum may suggest a low photospheric gravity, i.e$.$ a young age. However, this is not supported by the strong alkali lines (K\,{\sc i}, Na\,{\sc i}, Cs\,{\sc i}) of the optical spectrum. Using the optical spectroscopic data kindly provided by J$.$ Davy Kirkpatrick (as in McLean et al$.$ \cite{mclean03}), we have imposed an upper limit on the H$\alpha$ emission (Table~\ref{polarimetry3}), and have confirmed the brown dwarf nature of J2244$-$01 by detecting Li\,{\sc i} (Rebolo et al$.$ \cite{rebolo92}). It has been suggested (Golimowski et al$.$ \cite{golimowski04}, and references therein) that the observed spectroscopic and photometric properties of this brown dwarf can be attributed to heterogeneous, very thick cloud decks (that may cause an unusually strong veiling in the near-infrared wavelengths) resulting from a high metal abundance. Thick dust clouds would give rise to multiple scattering of photons, which in turn would yield less polarization than single scattering (Sengupta \cite{sengupta03}). Nevertheless, even the single scattering models of Sengupta \cite{sengupta03} show difficulties in explaining the significant amplitude of our polarimetric measurement. On the assumption that polarization is intrinsic to the object atmosphere and the presence of dust clouds, one possible way to account for the observed high polarization is by invoking rather large-size grains. In that case, J2244$+$20 should show polarization in the near-infrared wavelengths as well. Other possibility is related to the presence of a surrounding dusty disk or shell. As can be seen in the literature, many T\,Tauri stars show similar polarization degrees to that of J2244$+$20. Therefore, based on our data, we cannot discard that this dwarf is a young object. The determination of the astrometric parallax would help constrain its age. In addition, further photopolarimetric monitoring of this brown dwarf will be particularly interesting. 

\subsubsection{DENIS-P\,J225210.73$-$173013.4}

This is the coolest dwarf (L7.5V) in our sample. According to Kendall et al$.$ \cite{kendall04}, J2252$-$17 is probably located at 8.3\,pc (if it is a single object), being one of the nearest L dwarfs to the Sun. The averaged value of all our polarimetric measurements does not indicate detection because they differ by more than 5\,$\sigma$ the uncertainties. However, on the basis of our criterion, the data of one of the epochs support the presence of polarized photons in the $I$-band light from J2252$-$17. We note that only the averaged polarimetric value is plotted in Figs$.$~\ref{qupol} and~\ref{ipol}, and that the measurement with the highest $P/\sigma_P$ is shown in Table~\ref{polarimetry3}. The polarization degree of J2252$-$17 is similar to those of mid-L dwarfs, but is significantly smaller than the polarization degree of J2244$+$20 (L6.5V).

\section{Final remarks and conclusions}

We have conducted Johnson $I$-band (850\,nm) polarimetric observations of 35 L-type (L0--L7.5) field dwarfs and 10 M dwarfs (M4.5--M9.5). Johnson $R$-band (641\,nm) polarimetric data were also collected for two of the brightest targets in our sample: the L3.5 dwarf 2MASS\,J00361617$+$1821104, for which M\'enard et al$.$ \cite{menard02} have previously detected significant polarization, and the M7 dwarf CFHT-BD-Tau\,4, which is a brown dwarf member of the Taurus star-forming region. Because of the typical average uncertainty of our measurements ($P$), we can easily confirm polarization of objects with $P$\,$\ge$\,0.4\%. Eleven (10 L and 1 M) dwarfs show linear polarization degrees that comply with the following criterion: $P$/$\sigma_P$\,$\ge$\,3, where $\sigma_P$ is the error bar. For these, our measurements are in the interval $P$\,=\,0.2--2.5\%. The observed polarization is intrinsic to the objects. We have compared the fraction of polarized M and L dwarfs in our sample, and have found it to be higher for the cool L types. This is a clear evidence for the presence of considerable amounts of dust in ultracool atmospheres. 

We argue that the most viable origin of the observed polarization in our sample is photon scattering by heterogeneous dust clouds in a rotationally-induced oblate photosphere. In some cases, the possible presence of dusty disks or dust shells gives rise to comparatively high polariztion. The linear polarization degree of 2MASS\,J00361617$+$1821104 (L3.5, $\sim$1900\,K) appears to decrease with increasing wavelength (from 641 up to 850\,nm), suggesting that the grain growth lies in the submicron regime throughout the dwarf atmosphere. Our polarimetric data of the young brown dwarf CFHT-BD-Tau\,4 support the presence of a circum(sub)stellar dusty disk, which was previously suspected to exist from the observed near-infrared excesses and persistent, strong H$\alpha$ emission. 2MASS\,J22443167$+$2043433 (L6.5), which is a peculiar brown dwarf known for its very red colors and unusual near-infrared spectrum, shows the largest polarization degree in our sample ($P$\,=\,2.5\,$\pm$\,0.5\%). Rather large photospheric dust grains (possibly related to high metallicity) and/or the presence of a disk may account for such a considerable polarimetric amplitude.

We have compared our polarimetric measurements to $I$-band photometric varibility and rotation (using spectroscopic $v$\,sin\,$i$ values). No obvious correlation is seen between polarization and the projected rotational velocities. Three likely polarized L-type dwarfs spanning spectral types L2.5--L5 show $I$-band light curves with rather small amplitudes ($\le$10\,mmag) over a few hours of photometric monitoring. More light curves and further polarimetric studies are needed to prove (or discard) any possible relation between photometric variability and polarization.

\acknowledgements

We thank M$.$ Alises, F$.$ Hoyo, J$.$ Aceituno and U$.$ Thiele for their assistance with the observations at the 2.2\,m Calar Alto Telescope. We thank the referee (S$.$ Sengupta) for his very valuable comments. We also thank R$.$ Rebolo for his support at the time we started this work. We are indebted to B$.$ Goldman for sharing with us data prior to publication. This paper is based on observations collected at the Centro Astron\'omico Hispano Alem\'an (CAHA) at Calar Alto, operated jointly by the Max-Planck Institut f\"ur Astronomie and the Instituto de Astrof\'\i sica de Andaluc\'\i a (CSIC). This research has made use of the SIMBAD database, operated at CDS, Strasbourg, France, and of data products from the Two Micron All Sky Survey, which is a joint project of the University of Massachusetts and the Infrared Processing and Analysis Center/California Institute of Technology, funded by the National Aeronautics and Space Administration and the National Science Foundation. This work was partly carried out with support from the projects AYA2003-05355 and Ram\'on y Cajal. 

Facilities: \facility{CAHA 2.2m (CAFOS)}.

\clearpage

\clearpage


\begin{deluxetable}{lccccc}
\tablecolumns{6}
\tablewidth{0pc}
\tablecaption{Log of photopolarimetric observations.\label{observations}}
\tablehead{
\colhead{Object} & \colhead{$J$\tablenotemark{a}} & \colhead{UT Date\tablenotemark{b}} & \colhead{Exposure\tablenotemark{c}} & \colhead{Air mass} & \colhead{FWHM} \\
\colhead{      } & \colhead{                    } & \colhead{                        } & \colhead{(s)                      } & \colhead{        } & \colhead{(arcsec)} }
\startdata
BRI\,0021$-$0214 (DY\,Psc)  & 11.99 & 2003 Aug 30 & 2$\times$300,  2$\times$300,  0,            0            & 1.37 & 1.7 \\
                            &       & 2004 Aug 14 & 2$\times$40,   2$\times$40,   2$\times$40,  2$\times$40  & 1.50 & 1.2 \\
LP\,349$-$25                & 10.61 & 2004 Aug 13 & 2$\times$10,   2$\times$10,   2$\times$10,  2$\times$10  & 1.03 & 1.0 \\
2MASS\,J00361617$+$1821104  & 12.47 & 2003 Aug 29 & 2$\times$600,  2$\times$600,  0,            0            & 1.08 & 1.9 \\
                            &       & 2003 Aug 29\tablenotemark{d} & 1$\times$1000, 3$\times$1000, 0, 0      & 1.06 & 1.9 \\
                            &       & 2004 Aug 14 & 2$\times$120,  2$\times$120,  2$\times$120, 2$\times$120 & 1.16 & 1.0 \\
                            &       & 2004 Aug 14\tablenotemark{d} & 1$\times$180,1$\times$180, 1$\times$180 & 1.13 & 1.0 \\
                            &       & 2004 Aug 15 & 2$\times$120,  2$\times$120,  2$\times$120, 2$\times$120 & 1.06 & 1.4 \\
                            &       & 2004 Aug 18 & 2$\times$150,  2$\times$150,  2$\times$150, 2$\times$150 & 1.08 & 2.7 \\
                            &       & 2004 Aug 18\tablenotemark{d} & 1$\times$400,1$\times$400, 1$\times$400 & 1.06 & 2.4 \\
2MASS\,J00452143$+$1634446  & 13.06 & 2003 Aug 29 & 2$\times$600,  2$\times$600,  0,            0            & 1.13 & 2.0 \\
                            &       & 2004 Aug 18 & 2$\times$200,  2$\times$200,  2$\times$200, 2$\times$200 & 1.07 & 2.6 \\
2MASS\,J00584253$-$0651239  & 14.31 & 2003 Aug 30 & 2$\times$700,  1$\times$700,  0,            0            & 1.44 & 1.6 \\
LP\,647$-$13                & 11.69 & 2004 Aug 14 & 1$\times$15,   1$\times$15,   1$\times$15,  1$\times$15  & 1.49 & 1.2 \\
2MASS\,J01410321$+$1804502  & 13.88 & 2003 Aug 30 & 2$\times$700,  2$\times$700,  0,            0            & 1.06 & 1.3 \\
                            &       & 2004 Aug 18 & 2$\times$300,  2$\times$300,  2$\times$300, 2$\times$300 & 1.06 & 2.7 \\
                            &       & 2004 Aug 19 & 2$\times$300,  2$\times$300,  2$\times$300, 2$\times$300 & 1.23 & 1.7 \\
2MASS\,J01443536$-$0716142  & 14.19 & 2003 Aug 30 & 1$\times$600,  1$\times$600,  0,            0            & 1.41 & 1.6 \\
                            &       & 2003 Sep 02 & 2$\times$700,  2$\times$700,  0,            0            & 1.40 & 2.0 \\
                            &       & 2004 Aug 14 & 2$\times$200,  2$\times$200,  1$\times$200, 1$\times$200 & 1.40 & 1.1 \\
                            &       & 2004 Aug 19 & 2$\times$300,  2$\times$300,  2$\times$300, 2$\times$300 & 1.49 & 2.0 \\
DENIS-P\,J0205.4$-$1159     & 14.59 & 2004 Aug 19 & 1$\times$500,  1$\times$500,  1$\times$500, 1$\times$500 & 1.55 & 2.5 \\
CFHT-BD-Tau\,4              & 12.17 & 2003 Aug 29 & 1$\times$300,  1$\times$300,  0,            0            & 1.10 & 2.4 \\
                            &       & 2003 Aug 30\tablenotemark{d} & 2$\times$300,  1$\times$300,  0, 0      & 1.10 & 1.6 \\
                            &       & 2003 Aug 31 & 2$\times$300,  2$\times$300,  0,            0            & 1.15 & 2.2 \\
                            &       & 2003 Sep 02\tablenotemark{d} & 2$\times$300,  2$\times$300,  0, 0      & 1.10 & 2.3 \\
                            &       & 2003 Sep 02 & 2$\times$300,  2$\times$300,  0,            0            & 1.15 & 2.1 \\
2MASS\,J10452400$-$0149576  & 13.16 & 2004 May 17 & 2$\times$200,  2$\times$200,  1$\times$200, 1$\times$200 & 1.40 & 1.6 \\
DENIS-P\,J104842.8$+$011158 & 12.92 & 2004 May 16 & 2$\times$200,  2$\times$200,  2$\times$200, 2$\times$200 & 1.43 & 1.5 \\
2MASS\,J11083081$+$6830169  & 13.12 & 2004 May 20 & 2$\times$300,  2$\times$300,  2$\times$300, 2$\times$300 & 1.20 & 1.2 \\
Kelu 1                      & 13.41 & 2004 May 16 & 1$\times$360,  1$\times$360,  1$\times$360, 1$\times$360 & 2.25 & 2.2 \\
2MASS\,J14122449$+$1633115  & 13.89 & 2004 May 17 & 2$\times$300,  2$\times$300,  2$\times$300, 2$\times$300 & 1.07 & 2.5 \\
2MASS\,J14392836$+$1929149  & 12.76 & 2004 May 16 & 2$\times$180,  2$\times$180,  2$\times$180, 2$\times$180 & 1.05 & 1.3 \\
2MASS\,J15065441$+$1321060  & 13.36 & 2004 May 16 & 2$\times$300,  2$\times$300,  2$\times$300, 2$\times$300 & 1.10 & 1.5 \\
2MASS\,J15074769$-$1627386  & 12.83 & 2004 May 18 & 2$\times$300,  2$\times$300,  2$\times$300, 2$\times$300 & 1.70 & 2.8 \\
2MASS\,J15150083$+$4847416  & 14.11 & 2004 May 18 & 2$\times$300,  2$\times$300,  2$\times$300, 2$\times$300 & 1.06 & 2.2 \\
DENIS-P\,J153941.9$-$052042 & 13.92 & 2003 Aug 30 & 1$\times$900,  1$\times$900,  0,            0            & 2.00 & 2.1 \\
                            &       & 2004 May 17 & 2$\times$360,  2$\times$360,  2$\times$360, 2$\times$360 & 1.40 & 1.7 \\
2MASS\,J15525906$+$2948485  & 13.48 & 2004 May 20 & 2$\times$300,  2$\times$300,  2$\times$300, 2$\times$300 & 1.02 & 1.9 \\
2MASS\,J15551573$-$0956055  & 12.56 & 2004 May 19 & 2$\times$200,  2$\times$200,  2$\times$200, 2$\times$200 & 1.47 & 1.9 \\
                            &       & 2004 May 20 & 2$\times$200,  2$\times$200,  2$\times$200, 2$\times$200 & 1.65 & 2.1 \\
LSR\,J1610$-$0040           & 12.91 & 2004 May 19 & 2$\times$100,  2$\times$100,  2$\times$100, 2$\times$100 & 1.27 & 1.4 \\
2MASS\,J16154416$+$3559005  & 14.54 & 2004 May 19 & 2$\times$300,  2$\times$300,  2$\times$300, 2$\times$300 & 1.02 & 1.6 \\
                            &       & 2004 Aug 15 & 1$\times$600,  1$\times$600,  1$\times$600, 1$\times$600 & 1.09 & 1.6 \\
2MASS\,J16452211$-$1319516  & 12.45 & 2004 May 20 & 2$\times$200,  2$\times$200,  2$\times$200, 2$\times$200 & 1.75 & 2.1 \\
2MASS\,J16580380$+$7027015  & 13.29 & 2004 May 21 & 2$\times$300,  2$\times$300,  2$\times$300, 2$\times$300 & 1.22 & 1.4 \\
DENIS-P\,J170548.38$-$051645.7&13.31& 2004 Aug 13 & 2$\times$200,  2$\times$200,  1$\times$200, 1$\times$200 & 1.50 & 1.3 \\
                            &       & 2004 Aug 14 & 2$\times$400,  2$\times$400,  1$\times$400, 1$\times$400 & 1.70 & 1.5 \\
                            &       & 2004 Aug 16 & 2$\times$300,  2$\times$300,  2$\times$300, 2$\times$300 & 1.86 & 3.1 \\
2MASS\,J17073334$+$4301304  & 13.97 & 2004 May 17 & 1$\times$400,  1$\times$400,  1$\times$400, 1$\times$400 & 1.01 & 1.2 \\
                            &       & 2004 May 19 & 2$\times$300,  2$\times$300,  1$\times$300, 1$\times$300 & 1.03 & 1.6 \\
SDSS\,J171714.10$+$652622.2 & 14.95 & 2003 Aug 30 & 2$\times$900,  2$\times$900,  0,            0            & 1.22 & 2.0 \\
2MASS\,J17210390$+$3344160  & 13.62 & 2004 May 18 & 2$\times$300,  2$\times$300,  2$\times$300, 2$\times$300 & 1.02 & 1.6 \\
                            &       & 2004 Aug 15 & 1$\times$200,  1$\times$200,  1$\times$200, 1$\times$200 & 1.08 & 2.3 \\
LP\,44$-$162                & 11.45 & 2004 Aug 13 & 2$\times$20,   2$\times$20,   2$\times$20,  2$\times$20  & 1.20 & 1.0 \\  
G\,227$-$22                 &  8.54 & 2004 Aug 14 & 3$\times$5,    3$\times$5,    4$\times$5,   3$\times$5   & 1.13 & 1.0 \\ 
2MASS\,J18071593$+$5015316  & 12.93 & 2003 Aug 28 & 2$\times$1200, 2$\times$600,  0,            0            & 1.04 & 1.8 \\
                            &       & 2003 Aug 29 & 1$\times$300,  1$\times$300,  0,            0            & 1.03 & 1.6 \\
                            &       & 2004 May 17 & 2$\times$180,  2$\times$180,  2$\times$180, 2$\times$180 & 1.03 & 1.4 \\
                            &       & 2004 Aug 15 & 1$\times$150,  1$\times$150,  1$\times$150, 1$\times$150 & 1.12 & 2.4 \\
LSR\,J1835$+$3259           & 10.27 & 2004 May 19 & 2$\times$60,   2$\times$60,   2$\times$60,  2$\times$60  & 1.01 & 2.1 \\
LHS\,3406                   & 11.31 & 2004 Aug 12 & 2$\times$15,   2$\times$15,   2$\times$15,  2$\times$15  & 1.12 & 1.1 \\
SDSS\,J202820.32$+$005226.5 & 14.30 & 2003 Aug 28 & 2$\times$900,  2$\times$900,  0,            0            & 1.24 & 1.8 \\
                            &       & 2003 Aug 29 & 2$\times$700,  2$\times$700,  0,            0            & 1.30 & 1.6 \\
                            &       & 2004 Aug 13 & 2$\times$300,  2$\times$300,  2$\times$300, 2$\times$300 & 1.30 & 1.2 \\
                            &       & 2004 Aug 15 & 2$\times$300,  2$\times$300,  2$\times$300, 2$\times$300 & 1.26 & 1.9 \\
DENIS-P\,J205754.1$-$025229 & 13.12 & 2004 May 17 & 1$\times$180,  1$\times$180,  0,            0            & 1.50 & 1.3 \\
                            &       & 2004 May 19 & 2$\times$300,  2$\times$300,  0,            0            & 1.64 & 2.0 \\
                            &       & 2004 May 20 & 2$\times$300,  2$\times$200,  0,            0            & 1.46 & 1.8 \\
                            &       & 2004 Aug 14 & 1$\times$400,  1$\times$400,  0,            0            & 1.32 & 1.5 \\
                            &       & 2004 Aug 17 & 2$\times$200,  2$\times$200,  2$\times$200, 2$\times$200 & 1.35 & 1.6 \\
2MASS\,J21041491$-$1037369  & 13.84 & 2003 Aug 29 & 2$\times$700,  2$\times$700,  0,            0            & 1.50 & 1.6 \\
                            &       & 2004 Aug 16 & 2$\times$200,  2$\times$200,  2$\times$200, 2$\times$200 & 1.56 & 2.5 \\
2MASS\,J21580457$-$1550098  & 15.04 & 2003 Aug 29 & 2$\times$900,  2$\times$900,  0,            0            & 1.66 & 1.7 \\
                            &       & 2004 Aug 13 & 1$\times$1000, 1$\times$1000, 1$\times$1000,1$\times$1000& 1.67 & 1.2 \\
2MASS\,J22244381$-$0158521  & 14.07 & 2003 Aug 30 & 2$\times$600,  2$\times$600,  0,            0            & 1.32 & 2.3 \\
                            &       & 2004 Aug 15 & 2$\times$200,  2$\times$200,  2$\times$200, 2$\times$200 & 1.29 & 1.0 \\
                            &       & 2004 Aug 17 & 2$\times$300,  2$\times$300,  2$\times$300, 2$\times$300 & 2.06 & 2.9 \\
GJ\,4281                    & 10.77 & 2004 Aug 13 & 2$\times$20,   2$\times$20,   2$\times$20,  2$\times$20  & 1.94 & 1.6 \\
2MASS\,J22380742$+$4353179  & 13.84 & 2003 Aug 31 & 2$\times$900,  2$\times$900,  0,            0            & 1.00 & 1.7 \\
                            &       & 2004 Aug 16 & 2$\times$200,  2$\times$200,  2$\times$200, 2$\times$200 & 1.00 & 1.6 \\
2MASS\,J22443167$+$2043433  & 16.41 & 2003 Aug 28 & 1$\times$1800, 1$\times$1800, 0,            0            & 1.06 & 1.9 \\
                            &       & 2004 Aug 13 & 1$\times$600,  1$\times$600,  1$\times$600, 1$\times$600 & 1.09 & 1.0 \\
DENIS-P\,J225210.73$-$173013.4&14.31& 2004 Aug 13 & 1$\times$750,  1$\times$750,  1$\times$750, 1$\times$750 & 1.75 & 1.5 \\
                            &       & 2004 Aug 15 & 1$\times$900,  1$\times$900,  1$\times$900, 1$\times$900 & 1.74 & 1.4 \\
2MASS\,J23062928$-$0502285  & 11.35 & 2004 Aug 13 & 2$\times$20,   2$\times$20,   2$\times$20,  2$\times$20  & 1.36 & 1.0 \\
\enddata
\tablenotetext{a}{ 2MASS photometric data.}
\tablenotetext{b}{ All images were collected in the Johnson $I$-band, except when indicated.} 
\tablenotetext{c}{ Observations taken with the half-wave retarder positioned at 0, 45, 22.5 and 67.5\,deg, respectively.}
\tablenotetext{d}{ Images taken with the $R$-band filter.}
\end{deluxetable}

\clearpage

\begin{deluxetable}{lccccccc}
\tablecolumns{8}
\tablewidth{0pc}
\tablecaption{Linear photopolarimetric results.\label{polarimetry1}}
\tablehead{
\colhead{Object} & \colhead{Type} & \colhead{MJD} & \colhead{Filter} & \colhead{$Q/I$} & \colhead{$U/I$} & \colhead{$P$} & \colhead{$P/\sigma_P$\tablenotemark{a}} \\
\colhead{     } & \colhead{ }&\colhead{($-$50000)}& \colhead{      } & \colhead{(\%) } & \colhead{(\%) } &\colhead{(\%)} & \colhead{                             } }
\startdata
BRI\,0021      & M9.5V & 2881.0317 & $I$ & $+$0.11\,$\pm$\,0.18 & \nodata              & \nodata           & \nodata\\
               &       & 3231.0425 & $I$ & $-$0.15\,$\pm$\,0.26 & $+$0.17\,$\pm$\,0.11 & 0.23\,$\pm$\,0.28 & \nodata\\
LP\,349$-$25   & M8V   & 3230.1297 & $I$ & $+$0.20\,$\pm$\,0.17 & $+$0.17\,$\pm$\,0.32 & 0.27\,$\pm$\,0.37 & \nodata\\
J0036$+$18     & L3.5V & 2880.0636 & $I$ & $+$0.11\,$\pm$\,0.04 & \nodata              & \nodata           & \nodata\\
               &       & 2880.1065 & $R$ & $+$0.57\,$\pm$\,0.12 & \nodata              & \nodata           & (4.8)  \\
               &       & 3231.0590 & $I$ & $+$0.03\,$\pm$\,0.18 & $+$0.15\,$\pm$\,0.24 & 0.15\,$\pm$\,0.30 & \nodata\\
               &       & 3231.0736 & $R$ & $+$0.57\,$\pm$\,0.10 & $-$0.21\,$\pm$\,0.17 & 0.61\,$\pm$\,0.20 & 3.1    \\
               &       & 3232.1134 & $I$ & $+$0.02\,$\pm$\,0.13 & $+$0.03\,$\pm$\,0.12 & 0.04\,$\pm$\,0.18 & \nodata\\
               &       & 3235.0851 & $I$ & $-$0.05\,$\pm$\,0.15 & $-$0.04\,$\pm$\,0.15 & 0.06\,$\pm$\,0.21 & \nodata\\
               &       & 3235.1092 & $R$ & $+$0.67\,$\pm$\,0.19 & $+$0.47\,$\pm$\,0.12 & 0.82\,$\pm$\,0.23 & 3.6    \\
J0045$+$16     & L3.5V & 2880.1639 & $I$ & $-$0.01\,$\pm$\,0.13 & \nodata              & \nodata           & \nodata\\
               &       & 3235.1385 & $I$ & $-$0.17\,$\pm$\,0.04 & $-$0.04\,$\pm$\,0.10 & 0.18\,$\pm$\,0.10 & \nodata\\
J0058$-$06     & L0V   & 2881.0762 & $I$ & $-$0.18\,$\pm$\,0.08 & \nodata              & \nodata           & \nodata\\
LP\,647$-$13   & M9V   & 3231.0847 & $I$ & $+$0.27\,$\pm$\,0.12 & $+$0.16\,$\pm$\,0.12 & 0.31\,$\pm$\,0.18 & \nodata\\
J0141$+$18     & L4.5V & 2881.1282 & $I$ & $+$0.06\,$\pm$\,0.06 & \nodata              & \nodata           & \nodata\\
               &       & 3235.1809 & $I$ & $+$0.01\,$\pm$\,0.16 & $-$0.60\,$\pm$\,0.22 & 0.60\,$\pm$\,0.27 & \nodata\\
               &       & 3237.0689 & $I$ & $+$0.06\,$\pm$\,0.10 & $-$0.30\,$\pm$\,0.22 & 0.30\,$\pm$\,0.24 & \nodata\\
J0144$-$07     & L5V   & 2881.1616 & $I$ & $-$0.37\,$\pm$\,0.25 & \nodata              & \nodata           & \nodata\\
               &       & 2884.1328 & $I$ & $-$0.48\,$\pm$\,0.10 & \nodata              & \nodata           & (4.8)  \\
               &       & 3232.1839 & $I$ & $-$0.47\,$\pm$\,0.22 & $+$0.55\,$\pm$\,0.21 & 0.72\,$\pm$\,0.31 & \nodata\\
               &       & 3237.1155 & $I$ & $+$0.34\,$\pm$\,0.20 & $+$0.50\,$\pm$\,0.18 & 0.61\,$\pm$\,0.27 & \nodata\\
J0205$-$11     & L7V   & 3237.1586 & $I$ & $+$0.27\,$\pm$\,0.30 & $-$0.24\,$\pm$\,0.24 & 0.35\,$\pm$\,0.38 & \nodata\\
Tau\,4         & M7    & 2880.1947 & $I$ & $-$0.98\,$\pm$\,0.18 & \nodata              & \nodata           & (5.4)  \\
               &       & 2881.1902 & $R$ & $-$0.77\,$\pm$\,0.34 & \nodata              & \nodata           & \nodata\\
               &       & 2882.1703 & $I$ & $-$0.72\,$\pm$\,0.08 & \nodata              & \nodata           & (9.0)  \\
               &       & 2884.1643 & $R$ & $-$0.60\,$\pm$\,0.28 & \nodata              & \nodata           & \nodata\\
               &       & 2884.1822 & $I$ & $-$0.84\,$\pm$\,0.07 & \nodata              & \nodata           & (12.0) \\
J1045$-$01     & L1V   & 3142.8618 & $I$ & $+$0.07\,$\pm$\,0.50 & $+$0.06\,$\pm$\,0.10 & 0.10\,$\pm$\,0.51 & \nodata\\
J1048$+$01     & L1V   & 3141.8960 & $I$ & $-$0.01\,$\pm$\,0.02 & $+$0.07\,$\pm$\,0.09 & 0.07\,$\pm$\,0.10 & \nodata\\
J1108$+$68     & L1V   & 3145.8832 & $I$ & $-$0.03\,$\pm$\,0.03 & $+$0.02\,$\pm$\,0.04 & 0.04\,$\pm$\,0.05 & \nodata\\
Kelu 1         & L2.5V & 3141.9306 & $I$ & $+$0.79\,$\pm$\,0.23 & $+$0.13\,$\pm$\,0.13 & 0.80\,$\pm$\,0.27 & 3.0    \\
J1412$+$16     & L0.5V & 3142.9703 & $I$ & $+$0.29\,$\pm$\,0.13 & $+$0.49\,$\pm$\,0.13 & 0.57\,$\pm$\,0.19 & 3.0    \\
J1439$+$19     & L1V   & 3141.9583 & $I$ & $+$0.10\,$\pm$\,0.09 & $-$0.09\,$\pm$\,0.01 & 0.14\,$\pm$\,0.10 & \nodata\\
J1506$+$13     & L3V   & 3142.0115 & $I$ & $-$0.07\,$\pm$\,0.13 & $+$0.16\,$\pm$\,0.22 & 0.17\,$\pm$\,0.26 & \nodata\\
J1507$-$16     & L5V   & 3143.0168 & $I$ & $+$0.63\,$\pm$\,0.25 & $+$1.21\,$\pm$\,0.16 & 1.36\,$\pm$\,0.30 & 4.5    \\
J1515$+$48     & L6V   & 3143.0564 & $I$ & $+$0.33\,$\pm$\,0.04 & $+$0.21\,$\pm$\,0.30 & 0.39\,$\pm$\,0.30 & \nodata\\
J1539$-$05     & L4V   & 2881.8530 & $I$ & $-$0.39\,$\pm$\,0.39 & \nodata              & \nodata           & \nodata\\
               &       & 3142.0550 & $I$ & $+$0.07\,$\pm$\,0.16 & $-$0.07\,$\pm$\,0.01 & 0.10\,$\pm$\,0.16 & \nodata\\
J1552$+$29     & L1V   & 3145.0435 & $I$ & $-$0.63\,$\pm$\,0.10 & $+$0.18\,$\pm$\,0.28 & 0.66\,$\pm$\,0.29 & \nodata\\
J1555$-$09     & L1V   & 3144.0143 & $I$ & $-$0.16\,$\pm$\,0.20 & $+$0.10\,$\pm$\,0.31 & 0.18\,$\pm$\,0.37 & \nodata\\
               &       & 3145.0790 & $I$ & $-$0.03\,$\pm$\,0.34 & $+$0.12\,$\pm$\,0.05 & 0.12\,$\pm$\,0.34 & \nodata\\
J1610$-$00     & sdL   & 3144.0379 & $I$ & $-$0.07\,$\pm$\,0.15 & $-$0.09\,$\pm$\,0.14 & 0.11\,$\pm$\,0.21 & \nodata\\
J1615$+$35     & L3V   & 3144.0665 & $I$ & $+$0.04\,$\pm$\,0.02 & $-$0.28\,$\pm$\,0.20 & 0.29\,$\pm$\,0.21 & \nodata\\
               &       & 3232.8659 & $I$ & $-$0.25\,$\pm$\,0.15 & $+$0.07\,$\pm$\,0.10 & 0.25\,$\pm$\,0.18 & \nodata\\
J1645$-$13     & L1.5V & 3145.1090 & $I$ & $+$0.03\,$\pm$\,0.27 & $-$0.05\,$\pm$\,0.23 & 0.06\,$\pm$\,0.36 & \nodata\\
J1658$+$70     & L1V   & 3146.1180 & $I$ & $-$0.12\,$\pm$\,0.32 & $-$0.09\,$\pm$\,0.05 & 0.15\,$\pm$\,0.33 & \nodata\\
J1705$-$05\tablenotemark{b}& L4V   & 3230.8946 & $I$ & $-$0.12\,$\pm$\,0.10 & $+$0.05\,$\pm$\,0.10 & 0.13\,$\pm$\,0.14 & \nodata\\
               &       & 3231.9232 & $I$ & $-$0.12\,$\pm$\,0.08 & $-$0.12\,$\pm$\,0.14 & 0.17\,$\pm$\,0.17 & \nodata\\
               &       & 3233.9310 & $I$ & $-$0.17\,$\pm$\,0.28 & $-$0.09\,$\pm$\,0.88 & 0.19\,$\pm$\,0.92 & \nodata\\
J1707$+$43     & L0.5V & 3142.0975 & $I$ & $+$0.03\,$\pm$\,0.03 & $-$0.23\,$\pm$\,0.07 & 0.23\,$\pm$\,0.08 & \nodata\\
               &       & 3144.1020 & $I$ & $-$0.10\,$\pm$\,0.17 & $-$0.22\,$\pm$\,0.10 & 0.24\,$\pm$\,0.19 & \nodata\\
J1717$+$65     & L4V   & 2881.8928 & $I$ & $+$0.15\,$\pm$\,0.36 & \nodata              & \nodata           & \nodata\\
J1721$+$33     & L3V   & 3143.1101 & $I$ & $-$0.12\,$\pm$\,0.24 & $-$0.11\,$\pm$\,0.22 & 0.17\,$\pm$\,0.33 & \nodata\\
               &       & 3232.9018 & $I$ & $-$0.29\,$\pm$\,0.18 & $-$0.35\,$\pm$\,0.37 & 0.45\,$\pm$\,0.41 & \nodata\\
LP\,44$-$162   & M7.5V & 3230.8669 & $I$ & $-$0.17\,$\pm$\,0.09 & $+$0.03\,$\pm$\,0.05 & 0.18\,$\pm$\,0.10 & \nodata\\
G\,227$-$22    & M4.5V & 3231.8864 & $I$ & $+$0.27\,$\pm$\,0.23 & $+$0.01\,$\pm$\,0.24 & 0.27\,$\pm$\,0.34 & \nodata\\
J1807$+$50     & L1.5V & 2879.8711 & $I$ & $+$0.05\,$\pm$\,0.22 & \nodata              & \nodata           & \nodata\\
               &       & 2880.1065 & $I$ & $+$0.24\,$\pm$\,0.05 & \nodata              & \nodata           & (4.8)  \\ 
               &       & 3142.1374 & $I$ & $+$0.11\,$\pm$\,0.06 & $+$0.00\,$\pm$\,0.08 & 0.11\,$\pm$\,0.10 & \nodata\\
               &       & 3232.9553 & $I$ & $-$0.01\,$\pm$\,0.07 & $+$0.02\,$\pm$\,0.10 & 0.03\,$\pm$\,0.12 & \nodata\\
J1835$+$32     & M8.5V & 3144.1582 & $I$ & $+$0.02\,$\pm$\,0.03 & $+$0.04\,$\pm$\,0.01 & 0.05\,$\pm$\,0.03 & \nodata\\
LHS\,3406      & M5.5V & 3229.9890 & $I$ & $+$0.11\,$\pm$\,0.06 & $-$0.19\,$\pm$\,0.09 & 0.22\,$\pm$\,0.11 & \nodata\\
J2028$+$00     & L3V   & 2879.8554 & $I$ & $+$0.19\,$\pm$\,0.09 & \nodata              & \nodata           & \nodata\\
               &       & 2880.8796 & $I$ & $-$0.37\,$\pm$\,0.16 & \nodata              & \nodata           & \nodata\\
               &       & 3230.9235 & $I$ & $-$0.13\,$\pm$\,0.08 & $-$0.05\,$\pm$\,0.27 & 0.14\,$\pm$\,0.28 & \nodata\\
               &       & 3232.9848 & $I$ & $-$0.12\,$\pm$\,0.13 & $+$0.27\,$\pm$\,0.33 & 0.31\,$\pm$\,0.36 & \nodata\\
J2057$-$02     & L1.5V & 3142.1591 & $I$ & $-$0.23\,$\pm$\,0.08 & \nodata              & \nodata           & \nodata\\
               &       & 3144.1290 & $I$ & $+$0.17\,$\pm$\,0.11 & \nodata              & \nodata           & \nodata\\
               &       & 3145.1435 & $I$ & $+$0.19\,$\pm$\,0.26 & \nodata              & \nodata           & \nodata\\
               &       & 3231.9446 & $I$ & $+$0.07\,$\pm$\,0.11 & \nodata              & \nodata           & \nodata\\
               &       & 3234.0142 & $I$ & $+$0.33\,$\pm$\,0.15 & $-$0.03\,$\pm$\,0.37 & 0.33\,$\pm$\,0.39 & \nodata\\
J2104$-$10     & L2.5V & 2880.9339 & $I$ & $-$0.04\,$\pm$\,0.50 & \nodata              & \nodata           & \nodata\\
               &       & 3233.0264 & $I$ & $-$0.46\,$\pm$\,0.25 & $-$0.20\,$\pm$\,0.25 & 0.50\,$\pm$\,0.35 & \nodata\\
J2158$-$15     & L4V   & 2880.9939 & $I$ & $+$0.51\,$\pm$\,0.14 & \nodata              & \nodata           & (3.6)  \\
               &       & 3230.0249 & $I$ & $+$1.38\,$\pm$\,0.28 & $+$0.09\,$\pm$\,0.21 & 1.38\,$\pm$\,0.35 & 3.9    \\
J2224$-$01     & L4.5V & 2881.9704 & $I$ & $+$0.27\,$\pm$\,0.21 & \nodata              & \nodata           & \nodata\\
               &       & 3232.0365 & $I$ & $+$0.19\,$\pm$\,0.31 & $+$0.19\,$\pm$\,0.31 & 0.27\,$\pm$\,0.44 & \nodata\\
               &       & 3234.8934 & $I$ & $-$0.01\,$\pm$\,0.32 & $-$0.30\,$\pm$\,0.20 & 0.30\,$\pm$\,0.38 & \nodata\\
GJ\,4281       & M6.5V & 3230.9575 & $I$ & $-$0.07\,$\pm$\,0.13 & $-$0.04\,$\pm$\,0.07 & 0.08\,$\pm$\,0.15 & \nodata\\
J2238$+$43     & L1V   & 2882.0179 & $I$ & $+$0.01\,$\pm$\,0.07 & \nodata              & \nodata           & \nodata\\
               &       & 3233.0731 & $I$ & $-$0.14\,$\pm$\,0.28 & $-$0.20\,$\pm$\,0.18 & 0.24\,$\pm$\,0.34 & \nodata\\
J2244$+$20     & L6.5V & 2879.9894 & $I$ & $-$1.56\,$\pm$\,0.76 & \nodata              & \nodata           & \nodata\\
               &       & 3231.0068 & $I$ & $-$2.02\,$\pm$\,0.61 & $+$1.68\,$\pm$\,0.41 & 2.62\,$\pm$\,0.74 & 3.5    \\
J2252$-$17     & L7.5V & 3230.0896 & $I$ & $-$0.26\,$\pm$\,0.09 & $+$0.36\,$\pm$\,0.14 & 0.45\,$\pm$\,0.17 & \nodata\\
               &       & 3232.0758 & $I$ & $-$0.35\,$\pm$\,0.11 & $-$0.51\,$\pm$\,0.12 & 0.62\,$\pm$\,0.17 & 3.6    \\
J2306$-$05     & M7.5V & 3230.0586 & $I$ & $-$0.01\,$\pm$\,0.13 & $+$0.04\,$\pm$\,0.13 & 0.04\,$\pm$\,0.19 & \nodata\\
\enddata
\tablenotetext{a}{ Values in brackets are obtained from $Q/I$. We show values if $P/\sigma_P$\,$\ge$\,3.}
\tablenotetext{b}{ It is a double object with a separation less than 1.3\,arcsec.}
\end{deluxetable}

\clearpage

\begin{deluxetable}{lcccccccc}
\tablecolumns{7}
\tablewidth{0pc}
\tablecaption{Averages of linear polarimetric measurements.\label{polarimetry2}}
\tablehead{
\colhead{Object} & \colhead{Type} & \colhead{Filter} & \colhead{$Q/I$} & \colhead{$N$\tablenotemark{a}} & \colhead{$U/I$} & \colhead{$N$\tablenotemark{b}} & \colhead{$P$} & \colhead{$P/\sigma_P$\tablenotemark{c}} \\
\colhead{      } & \colhead{    } & \colhead{      } & \colhead{(\%) } & \colhead{   } & \colhead{(\%) } & \colhead{   } & \colhead{(\%)}& \colhead{                             } }
\startdata
BRI\,0021      & M9.5V & $I$ & $-$0.03\,$\pm$\,0.12 & 2 & $+$0.17\,$\pm$\,0.11 & 1 & 0.18\,$\pm$\,0.17 & \nodata \\
J0045$+$16     & L3.5V & $I$ & $-$0.09\,$\pm$\,0.08 & 2 & $-$0.04\,$\pm$\,0.10 & 1 & 0.10\,$\pm$\,0.12 & \nodata \\
J0036$+$18     & L3.5V & $I$ & $+$0.02\,$\pm$\,0.03 & 4 & $+$0.05\,$\pm$\,0.05 & 3 & 0.05\,$\pm$\,0.06 & \nodata \\
               &       & $R$ & $+$0.60\,$\pm$\,0.03 & 3 & $+$0.13\,$\pm$\,0.33 & 2 & 0.62\,$\pm$\,0.33 & \nodata \\
J0141$+$18     & L4.5V & $I$ & $+$0.04\,$\pm$\,0.01 & 3 & $-$0.45\,$\pm$\,0.15 & 2 & 0.45\,$\pm$\,0.15 & 3.0     \\
J0144$-$07     & L5V   & $I$ & $-$0.24\,$\pm$\,0.19 & 4 & $+$0.53\,$\pm$\,0.02 & 2 & 0.58\,$\pm$\,0.19 & 3.0     \\
Tau\,4         & M7    & $I$ & $-$0.85\,$\pm$\,0.07 & 3 & \nodata              &\nodata& \nodata       & (12.0)  \\
               &       & $R$ & $-$0.69\,$\pm$\,0.09 & 2 & \nodata              &\nodata& \nodata       & (7.7)   \\
J1555$-$09     & L1V   & $I$ & $-$0.09\,$\pm$\,0.09 & 2 & $+$0.11\,$\pm$\,0.01 & 2 & 0.14\,$\pm$\,0.09 & \nodata \\
J1615$+$35     & L3V   & $I$ & $-$0.11\,$\pm$\,0.14 & 2 & $-$0.11\,$\pm$\,0.17 & 2 & 0.15\,$\pm$\,0.22 & \nodata \\
J1705$-$05     & L4V   & $I$ & $-$0.14\,$\pm$\,0.02 & 3 & $-$0.05\,$\pm$\,0.08 & 3 & 0.15\,$\pm$\,0.09 & \nodata \\
J1707$+$43     & L0.5V & $I$ & $-$0.03\,$\pm$\,0.06 & 2 & $-$0.22\,$\pm$\,0.01 & 2 & 0.23\,$\pm$\,0.06 & 3.8     \\
J1721$+$33     & L3V   & $I$ & $-$0.21\,$\pm$\,0.08 & 2 & $-$0.23\,$\pm$\,0.11 & 2 & 0.31\,$\pm$\,0.14 & \nodata \\
J1807$+$50     & L1.5V & $I$ & $+$0.09\,$\pm$\,0.05 & 4 & $+$0.01\,$\pm$\,0.01 & 2 & 0.10\,$\pm$\,0.06 & \nodata \\
J2028$+$00     & L3V   & $I$ & $-$0.11\,$\pm$\,0.11 & 4 & $+$0.11\,$\pm$\,0.16 & 2 & 0.16\,$\pm$\,0.20 & \nodata \\
J2057$-$02     & L1.5V & $I$ & $+$0.11\,$\pm$\,0.09 & 5 & $-$0.03\,$\pm$\,0.37 & 1 & 0.11\,$\pm$\,0.38 & \nodata \\
J2104$-$10     & L2.5V & $I$ & $-$0.25\,$\pm$\,0.21 & 2 & $-$0.20\,$\pm$\,0.25 & 1 & 0.32\,$\pm$\,0.33 & \nodata \\
J2158$-$15     & L4V   & $I$ & $+$0.94\,$\pm$\,0.43 & 2 & $+$0.09\,$\pm$\,0.21 & 1 & 0.95\,$\pm$\,0.48 & \nodata \\
J2224$-$01     & L4.5V & $I$ & $+$0.15\,$\pm$\,0.08 & 3 & $-$0.05\,$\pm$\,0.24 & 2 & 0.16\,$\pm$\,0.26 & \nodata \\
J2238$+$43     & L1V   & $I$ & $-$0.06\,$\pm$\,0.07 & 2 & $-$0.20\,$\pm$\,0.19 & 1 & 0.21\,$\pm$\,0.20 & \nodata \\
J2244$+$20     & L6.5V & $I$ & $-$1.79\,$\pm$\,0.22 & 2 & $+$1.68\,$\pm$\,0.41 & 1 & 2.45\,$\pm$\,0.47 & 5.2     \\
J2252$-$17     & L7.5V & $I$ & $-$0.31\,$\pm$\,0.04 & 2 & $-$0.08\,$\pm$\,0.43 & 2 & 0.32\,$\pm$\,0.43 & \nodata \\
\enddata
\tablenotetext{a}{ Number of $Q/I$ measurements.}
\tablenotetext{b}{ Number of $U/I$ measurements.}
\tablenotetext{c}{ Values in brackets are obtained from $Q/I$.}
\end{deluxetable}

\clearpage

\begin{deluxetable}{lcccrccccc}
\tablecolumns{10}
\tablewidth{0pc}
\tablecaption{Likely and possible polarized dwarfs.\label{polarimetry3}}
\tablehead{
\colhead{Object} & \colhead{Type} & \colhead{Filter} & \colhead{$P$} & \colhead{$\theta$} & \colhead{$P/\sigma_P$\tablenotemark{a}} & \colhead{H$\alpha$\tablenotemark{b}} & \colhead{Li\,{\sc i}\tablenotemark{b}} & \colhead{$v$\,sin\,$i$} & \colhead{IR excess}\\
\colhead{}       & \colhead{}     & \colhead{}       &\colhead{(\%)} & \colhead{(deg)}    & \colhead{}                              & \colhead{(\AA)         } & \colhead{(\AA)           } & \colhead{(km\,s$^{-1}$)} & \colhead{             }} 
\startdata
J0036$+$18     & L3.5V & $R$ & 0.61\,$\pm$\,0.20 & 167\,$\pm$\,9  & 3.1   & $\le$0.5   & $\le$0.5 & 15      & no      \\
               &       & $R$ & 0.82\,$\pm$\,0.23 &  14\,$\pm$\,8  & 3.6   & $\le$0.5   & $\le$0.5 & 15      & no      \\
J0141$+$18     & L4.5V & $I$ & 0.45\,$\pm$\,0.15 & 135\,$\pm$\,2  & 3.0   & \nodata    & \nodata  & \nodata & \nodata \\
J0144$-$07     & L5V   & $I$ & 0.58\,$\pm$\,0.19 &  54\,$\pm$\,9  & 3.0   & $\le$3--23 & \nodata  & \nodata & \nodata \\
Tau\,4         & M7    & $I$ & $\ge$0.8          & \nodata        & (12.0)& 69--340    & \nodata  & 11:     & yes    \\
               &       & $R$ & $\ge$0.6          & \nodata        & (7.7) & 69--340    & \nodata  & 11:     & yes    \\
Kelu\,1        & L2.5V & $I$ & 0.80\,$\pm$\,0.27 &   3\,$\pm$\,6  & 3.0   & $\le$1--5  & 0.6--4.7 & 60      & no      \\
J1412$+$16     & L0.5V & $I$ & 0.57\,$\pm$\,0.19 &  25\,$\pm$\,8  & 3.0   & 4          & $\le$0.5 & 16.4    & \nodata \\
J1507$-$16     & L5V   & $I$ & 1.36\,$\pm$\,0.30 &  27\,$\pm$\,12 & 4.5   & $\le$0.5   & $\le$0.5 & 27.2    & no      \\
J1707$+$43     & L0.5V & $I$ & 0.23\,$\pm$\,0.06 & 128\,$\pm$\,8  & 3.8   & 35         & \nodata  & \nodata & \nodata \\
J2158$-$15     & L4V   & $I$ & 1.38\,$\pm$\,0.35 & 179\,$\pm$\,6  & 3.9   & \nodata    & \nodata  & \nodata & \nodata \\
J2244$+$20     & L6.5V & $I$ & 2.45\,$\pm$\,0.47 &  68\,$\pm$\,7  & 5.2   & $\le$5\tablenotemark{c} & 6\,$\pm$\,3\tablenotemark{c} & \nodata & very red\tablenotemark{d} \\
J2252$-$17     & L7.5V & $I$ & 0.62\,$\pm$\,0.16 & 115\,$\pm$\,8  & 3.9   & \nodata    & \nodata  & \nodata & \nodata \\
\enddata
\tablenotetext{a}{ Values in brackets are obtained from $Q/I$.}
\tablenotetext{b}{ Pseudo-equivalent widths. If detected, H$\alpha$ is seen in emission and the Li\,{\sc i} line at 670.8\,nm is seen in absorption.}
\tablenotetext{c}{ Measurements obtained from the Keck low-resolution spectrum provided by J$.$ Davy Kirkpatrick.}
\tablenotetext{d}{ The near-infrared and infrared colors of J2244$+$20 are significantly redder than those of other mid-L to late-L dwarfs.}
\end{deluxetable}

\clearpage

\begin{deluxetable}{lccc}
\tablecolumns{3}
\tablewidth{0pc}
\tablecaption{Frequency of polarized ultracool dwarfs.\label{freq}}
\tablehead{
\colhead{                               } & \colhead{M4.5V--M9.5V} & \colhead{L0V--L3V} & \colhead{L3.5V--L8V} }
\startdata
Our data\tablenotemark{a}                 & 0\%                    &     15\,$\pm$\,9\% & 43\,$\pm$\,17\% \\
Data of M\'enard et al$.$ \cite{menard02} & \nodata                &     25\%           & 50\%     \\
\enddata
\tablenotetext{a}{ For $P$ ($I$-band)\,$\ge$\,0.2\%. These values may be affected by some biases (see text).}
\end{deluxetable}

\clearpage

\begin{deluxetable}{lcccc}
\tablecolumns{5}
\tablewidth{0pc}
\tablecaption{Photometric variability of 18 dwarfs in our sample.\label{photvar}}
\tablehead{
\colhead{Object} & \colhead{$P$ ($I$-band)} & \colhead{$P/\sigma_P$} & \colhead{Phot$.$ var$.$} & \colhead{Ref$.$\tablenotemark{a}} }
\startdata
BRI\,0021   &  0.18\,$\pm$\,0.17 & \nodata &  yes     ($I$)        &  1   \\
J0036$+$18  &  0.05\,$\pm$\,0.06 & \nodata &  no ($I$), no ($JK_s$)&  2,3 \\
            &  0.20\,$\pm$\,0.03\tablenotemark{b} & 6.7\tablenotemark{b} & no ($I$), no ($JK_s$) &  2,3 \\
J0058$-$06  &  \nodata           & \nodata &  weak    ($I$)        &  2   \\
J0205$-$11  &  0.35\,$\pm$\,0.38 & \nodata &  no      ($K$)        &  4   \\
J1045$-$01  &  0.10\,$\pm$\,0.51 & \nodata &  no      ($I$)        &  5   \\
J1048$+$01  &  0.07\,$\pm$\,0.10 & \nodata &  yes     ($I$)        &  5   \\
J1108$+$68  &  0.04\,$\pm$\,0.05 & \nodata &  yes ($I$), yes ($I$) &  2,6 \\
Kelu\,1     &  0.80\,$\pm$\,0.27 & 3.0     &  yes ($I$)            &  7   \\
J1412$+$16  &  0.57\,$\pm$\,0.19 & 3.0     &  no      ($I$)        &  2   \\ 
J1439$+$19  &  0.14\,$\pm$\,0.10 & \nodata &  weak ($I$), no ($I$) &  2,8 \\
J1506$+$13  &  0.17\,$\pm$\,0.26 & \nodata &  no      ($I$)        &  2   \\
J1507$-$16  &  1.36\,$\pm$\,0.30 & 4.5     &  no ($I$), yes ($I$)  &  5,9 \\
J1615$+$35  &  0.15\,$\pm$\,0.22 & \nodata &  weak    ($I$)        &  2   \\
J1645$-$13  &  0.06\,$\pm$\,0.36 & \nodata &  no      ($I$)        &  5   \\
J1658$+$70  &  0.15\,$\pm$\,0.33 & \nodata &  yes     ($I$)        &  2   \\
J2028$+$00  &  0.16\,$\pm$\,0.20 & \nodata &  no      ($I$)        &  5   \\
J2057$-$02  &  0.11\,$\pm$\,0.38 & \nodata &  weak    ($I$)        &  5   \\
J2224$-$01  &  0.16\,$\pm$\,0.26 & \nodata &  yes ($I$), no ($JK_s$)& 2,3 \\
\enddata
\tablenotetext{a}{ References: 1 --- Mart\'\i n et al$.$ \cite{martin01a}; 2 --- Gelino et al$.$ \cite{gelino02}; 3 --- Caballero et al$.$ \cite{caballero03}; 4 --- Enoch et al$.$ \cite{enoch03}; 5 --- Koen \cite{koen03}; 6 --- Clarke et al$.$ \cite{clarke02b}; 7 --- Clarke et al$.$ \cite{clarke02a}; 8 --- Bailer-Jones \& Mundt \cite{bailer01}; 9 --- Goldman (2004, priv$.$ communication).}
\tablenotetext{b}{ From M\'enard et al$.$ \cite{menard02}.}
\end{deluxetable}

\end{document}